\documentclass{ar-1col}
\usepackage[numbers]{natbib}
\usepackage[]{amsmath,amssymb,amsthm}
\usepackage{url}
\newcommand{\mb}[1]{\mbox{\boldmath$#1$}}
\newcommand{\argmax}{\operatornamewithlimits{argmax}}
\newcommand{\argmin}{\operatornamewithlimits{argmin}}

\begin{document}

\markboth{Amit Singer and Fred J.~Sigworth}{Computational Methods for Single-Particle Cryo-EM}

\title{Computational Methods for Single-Particle Cryo-EM}

\author{Amit Singer$^1$ and Fred J. Sigworth$^2$
\affil{$^1$Department of Mathematics and Program in Applied and Computational Mathematics, Princeton University, Princeton, NJ 08544, USA; email: amits@math.princeton.edu}
\affil{$^2$Departments of Cellular and Molecular Physiology, Biomedical Engineering, and Molecular Biophysics and Biochemistry, Yale University, New Haven, CT 06520, USA; email: fred.sigworth@yale.edu}
}

\begin{abstract}
Single-particle electron cryomicroscopy (cryo-EM) is an increasingly popular technique for elucidating the three-dimensional structure of proteins and other biologically significant complexes at near-atomic resolution. It is an imaging method that does not require crystallization and can capture molecules in their native states.

In single-particle cryo-EM, the three-dimensional molecular structure needs to be determined from many noisy two-dimensional tomographic projections of individual molecules, whose orientations and positions are unknown. The high level of noise and the unknown pose parameters are two key elements that make reconstruction a challenging computational problem.  Even more challenging is the inference of structural variability and flexible motions when the individual molecules being imaged are in different conformational states.

This review discusses computational methods for structure determination by single-particle cryo-EM and their guiding principles from statistical inference, machine learning, and signal processing that also play a significant role in many other data science applications.

\end{abstract}

\maketitle

\tableofcontents

\section{INTRODUCTION}

In the past few years, electron microscopy of cryogenically-cooled specimens (cryo-EM) has become a major technique in structural biology \cite{nogales2016development}. Electrons interact strongly with matter, and they can be focused by magnetic lenses to yield magnified images in which individual macromolecules are visible \cite{henderson1995potential}. Noise obscures high-resolution details in such micrographs, so information must be combined and averaged from images of multiple copies of the macromolecules \cite{frank}.

There are two main approaches in cryo-EM. In electron tomography multiple images are taken from the same field of a frozen specimen as it is tilted to various angles. A 3-D tomogram is constructed, from which are excised small 3-D subtomograms, each representing one macromolecule. The subtomograms are averaged to yield a 3-D structure. The other approach, single-particle reconstruction (SPR), starts with a single image taken from each field of the specimen. In the field are randomly-oriented and positioned macromolecules, termed particles. Through subsequent image analysis the orientations of the particles are determined. Given the orientations, particle images can be grouped and averaged, and subsequent tomographic reconstruction yields a three-dimensional density map. SPR is the subject of this review.

A very important advantage of Cryo-EM SPR is that ordered arrays (crystals) of macromolecules are not required to allow the averaging of particle images. Sets of $10^5$ to $10^6$ particle images are typically used to obtain maps with near-atomic resolution, and in favorable cases datasets of this size can be acquired in a few hours of microscope time.

Because in SPR crystallization of the protein is not required, structure determination is greatly simplified especially for membrane proteins for which crystallization has been extremely difficult. This advantage has led to a surge in the number of protein structures deposited in the Protein Data Bank. This number has nearly doubled each year for the past five years. In the past year some 1300 cryo-EM structures were determined, representing about 10\% of all depositions (see \url{http://www.ebi.ac.uk/pdbe/emdb/}).

The process of extracting information from the images for SPR structure determination is completely computational. This review focuses on the primary computational methods and their underlying theory, as well as on important remaining computational challenges. The computational methods are implemented in one form or another in a variety of software packages that keep evolving. A list EM software is available at \url{https://www.emdataresource.org/emsoftware.html}. This review does not attempt to be a user manual and deliberately avoids mentioning specific software. Another useful online source closely related to this review is \url{http://3demmethods.i2pc.es} where readers can find a more complete list of publications related to computational methods.

\subsection{Overview of Structural Biology}

Molecular machines perform functions in living cells that include copying the DNA, transducing chemical and electrical signals, catalyzing reactions, causing mechanical movements, and transporting molecules into and out of cells. Molecular machines are macromolecules or macromolecular assemblies that can contain hundreds of thousands of atoms, and the goal of structural biology is to determine their three-dimensional structure--that is, the position of each atom in the machine--to try to understand the way in which the machine works. Machines such as a molecular motor or a solute transporter pass through various distinct conformations in the course of their reaction cycles. Thus a complete understanding of its operation requires the determination of multiple structures, one for each conformation.

Macromolecules are composed of polymers of small component parts. Nucleic acids (DNA and RNA) are composed of individual nucleotides, each conveying one of the four ``letters'' of the genetic code. Proteins are polymers of amino acids, of which there are 20 varieties. Coding sequences of DNA dictate the particular order of amino acids in the protein's polypeptide chain, and this sequence specifies in turn the particular three-dimensional shape into which the chain folds. Because of hydrogen-bonding patterns, portions of the chain often curl up into alpha helices or lie in extended, parallel strands of beta sheets. Nevertheless, the folding process has many degrees of freedom and may also be dependent on the presence of other macromolecules such as molecular chaperones. Thus, the prediction of the 3D structure of a protein from its sequence--called the ``protein folding problem''--has been successful only for some small proteins. The experimental determination of 3D structures of macromolecules is therefore a key step in understanding the fundamental processes in cells.

The size of a macromolecule is typically described as its molecular mass (or ``weight'') in daltons, where 1 Da is about the mass of a hydrogen atom. A typical amino acid residue is about 120 Da in mass; a large molecular machine might be a complex of multiple subunits totalling 1MDa in molecular weight, and have physical dimensions on the order of 200 angstroms.

The atomic structure of a macromolecule is obtained by fitting an atomic model into a 3D density map. The resolution of such a map is typically specified by the period of the highest-frequency Fourier component that can be faithfully estimated. The resolution needs to be about $1.5 \rm{\AA}$ if individual atoms of carbon, nitrogen etc. are to be located directly; however a map of $3-4 \rm{\AA}$ resolution is sufficient to produce an informative atomic model. Chemical constraints allow such lower-resolution, ``near-atomic resolution'' maps to be interpreted. Constraints come from the prior knowledge that proteins are composed of polypeptide chains, that the order of the amino-acid subunits in a chain is determined by the corresponding DNA sequence, and from knowledge of the geometric constraints of covalent bonds on the local arrangements of atoms.

\subsection{X-ray crystallography, NMR, and cryo-EM}

The first and most widely used method for determining macromolecular structures is X-ray crystallography. Development started in the 1960s, when it was found that macromolecules, especially small proteins, can be coaxed into forming highly-ordered crystals. When exposed to an X-ray beam and rotated through many angles, a crystal produces diffracted beams that are recorded as spots on film or an electronic detector. X-ray wavelengths near $1 \rm{\AA}$ are used. More and better data can be collected from crystals that are cooled to cryogenic temperatures, where effects of radiation damage to the crystal are mitigated. Radiation damage limits the allowable X-ray dose. The recorded spot intensities represent the magnitude-squared coefficients of a Fourier-series expansion of the 3D density of a single unit cell of the crystal. The phases of the coefficients can be determined by various other means, so that solving an X-ray structure is, in principle, simply the inversion of a discrete Fourier transform. The most difficult part of obtaining a structure is the process of purifying and crystallizing a purified sample of the macromolecule in question.

A second method for protein structure determination is based on nuclear magnetic resonance (NMR). The spin of an atomic nucleus produces a small magnetic dipole. In the presence of a strong background magnetic field the nuclear dipoles can be flipped to a higher-energy state by exposure to a radio-frequency (hundreds of megahertz) field. In decaying to the ground state, there is a distance-dependent probability of exciting nearby nuclear dipoles, and it is the measurement of this coupling that provides an estimate of the pairwise distance between atoms. In small proteins (fewer than about 200 amino acid residues) it is possible to distinguish the signals from individual residues and thereby assign distances to many specific atom pairs in the protein. The 3D structure is then obtained computationally by determining the global structure from the local distance estimates.

Cryo-EM is transmission microscopy using electrons, which like X-rays allow angstrom-scale features to be detected.  Also like X-ray crystallography, radiation damage is a limitation, even when mitigated by keeping the sample at low temperature. As it turns out, the amount of damage per useful scattering event is orders of magnitude smaller with electrons, which allows more information to be obtained from each copy of a macromolecular particle. Because of this, cryo-EM allows structures to be determined from relatively few particles, theoretically as few as $10^4$ \cite{henderson1995potential,glaeser1999electron}, as compared to $\sim 10^8$ macromolecules in a minimal-sized crystal \cite{holton2010minimum} for X-ray structure determination at a similar resolution.

A striking advantage of single-particle cryo-EM is its ability to work with heterogeneous ensembles of particles. In the same way that image analysis allows orientations of particles to be estimated, it also allows, within limits, the discrimination of distinct classes of particles. It is therefore possible to obtain structures of multiple conformations of a macromolecular machine from a single dataset. In some cases it is also possible to obtain 3D density maps of multiple proteins in a mixture.

\subsection{Example of a cryo-EM study}

\begin{figure}
\includegraphics[width=1.0\textwidth]{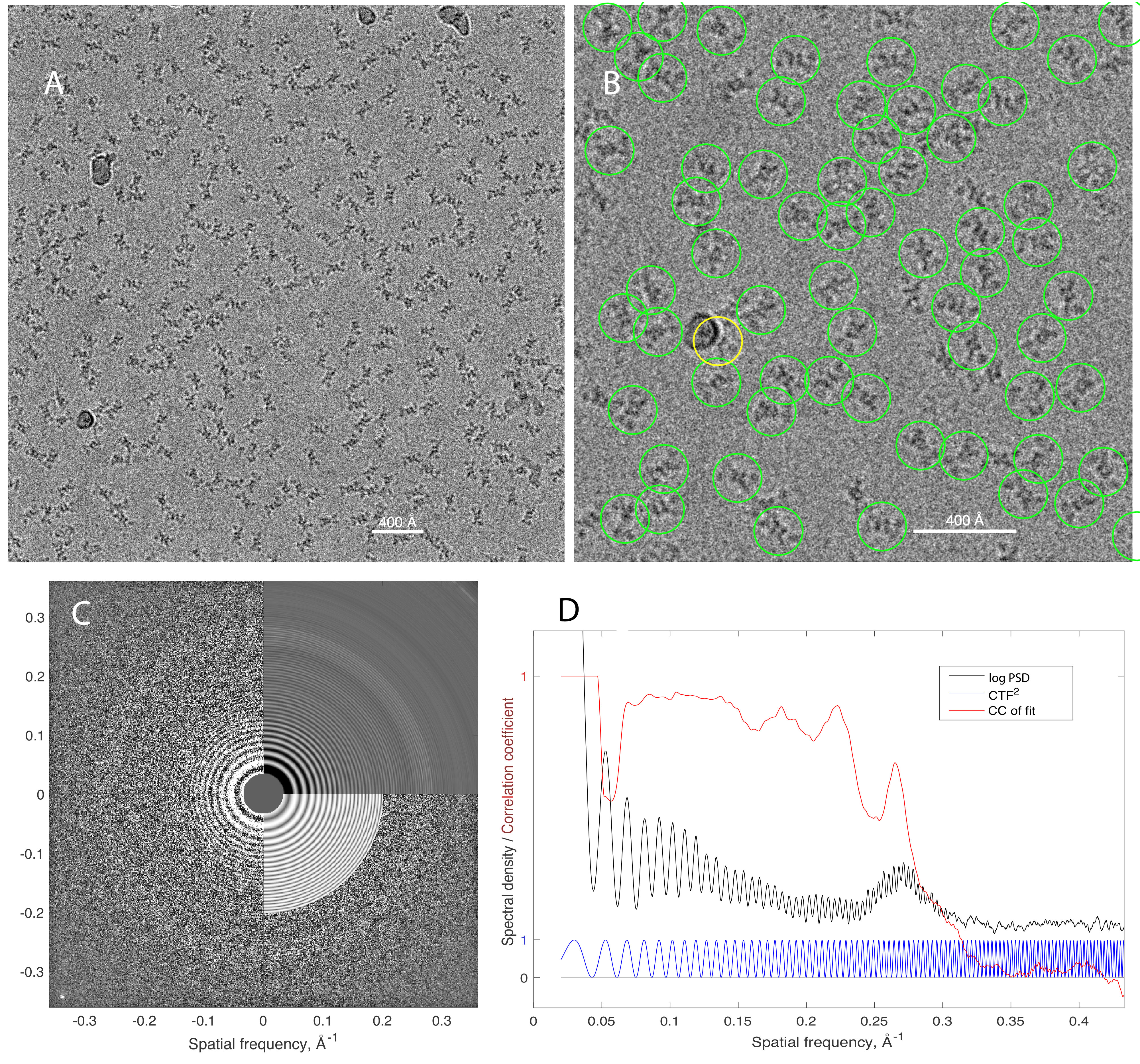}
\caption{A cryo-EM micrograph of PaaZ protein particles. {\bf A}, the entire micrograph, depicting a specimen area of $434 \times 434$ nm with a pixel size of $1.06 \rm{\AA}$. The field contains about 220 PaaZ particles along with a few contaminating ``ice balls''. The image was obtained on a Titan Krios microscope at 300kV, and is displayed after low-pass filtering for clarity. {\bf B}, a close-up of the micrograph with particles, selected automatically by template-matching, indicated by circles. One particle image selected in error (yellow circle) has an ice ball superimposed. {\bf C}, Fit of the CTF to the 2D power spectrum of the micrograph. The left half of the image shows the experimental power spectrum of the micrograph; the upper right quadrant shows the circularly-averaged power spectrum; the lower right quadrant shows the fitted ${\rm{CTF}}^2(|k|)$. The fitted parameters included defocus $= 2.66 \rm{\mu m}$ and astigmatism 49 nm. The bright spot in the lower-left corner comes from a $2.1 \rm{\AA}$ periodicity of the underlying crystalline graphene oxide substrate.  {\bf D}, a 1D representation of the power spectrum (black curve), obtained, after rough prewhitening, by approximately-circular averaging that takes into account astigmatism. Also shown is the theoretical contrast-transfer function $H^2$ (Eq. \ref{eq:ctf}; blue curve) and a plot of the cycle-by-cycle correlation coefficient between the two (red curve) showing that information is present for spatial frequencies $|k| \lesssim 0.3 \rm{\AA}^{-1}$.  The broad peak near $|k|=0.27 \rm{\AA^{-1}}$ comes from the amorphous ice itself; although not in a crystal lattice, water molecules have correlated positions with a nearest-neighbor distance of about $3.7 \rm{\AA}$. Experimental data in this and subsequent figures are from \cite{sathyanarayanan2019molecular}.}
\label{fig1}
\end{figure}

Throughout this review we use a recent cryo-EM study by Sathyanarayanan et al.\cite{sathyanarayanan2019molecular} to illustrate the steps in structure determination by cryo-EM. The object was the bacterial enzyme PaaZ which is important in the breakdown of environmental pollutants. PaaZ has two catalytic activities, a hydratase and a dehydrogenase, each having a distinct active site. Very efficient transfer of the small substrate molecule from one active site to the other had been measured, and now with the solved structure it is clear that the substrate pivots in a short ``tunnel'' connecting the active sites. The PaaZ complex is composed of six identical polypeptide chains totaling 500kDa in mass.

Cryo-EM specimens typically consist of a film of solution 200-600$\rm{\AA}$ in thickness,  containing the macromolecules of interest and frozen very rapidly by plunging into liquid ethane. The specimen is kept at cryogenic temperature throughout storage and imaging. In the PaaZ study the film of solution was formed over a graphene oxide substrate. A micrograph (\textbf{Figure \ref{fig1}}) taken with a high-end microscope and electron-counting camera, shows hundreds of dark particles on a noisy background. From 502 such micrographs 118,000 particle images were extracted, of which 86,400 were used in computing the final 3D density map. This map has $2.9 \rm{\AA}$ resolution and was used to build an atomic model of the protein complex, illustrated at the end of this article.

\section{IMAGE FORMATION MODEL}

\subsection{The electron microscope}
The first electron microscope was built in 1933 by Ernst Ruska, an invention for which he won the Nobel Prize in Physics in 1986. The electron microscope is based on the fundamental quantum mechanical principle of the wave-particle duality of electrons. The wavelength $\lambda$ of electrons with velocity $v$ is $\lambda = \frac{h}{m_ev}$, where $h$ is Planck's constant and $m_e$ is the relativistic electron mass.
Modern microscopes accelerate electrons to energies between 100 to 300 keV for which the associated wavelengths range from approximately $0.037~\rm{\AA}$ to $0.020~\rm{\AA}$.
Although, theoretically, sub-angstrom resolutions could be obtained, in practice, the resolution is limited to about $1~\rm{\AA}$ due to  aberrations in the magnetic objective lens.

For the purpose of imaging, the molecules are frozen in fixed positions. The solution of molecules is cooled so rapidly that water molecules do not have time to crystallize, forming a vitreous ice layer \cite{dubochet1982electron}. Rapid cooling prevents the formation of crystalline ice which would damage the macromolecules. The high contrast contribution of crystalline ice to the image would also overwhelm the faint signal coming from the molecules. Another important aspect of rapid cooling is that molecules do not have sufficient time to deform their structure from that in room temperature.

Images are formed by exposing the specimen to the electron beam. A majority of electrons recorded by the detector are those that pass through without any interaction, thus leading only to shot noise. The remaining electrons are scattered, either elastically or inelastically. Inelastic scattering damages the molecules (ionization, etc.), a process often referred to as radiation damage. Radiation damage imposes severe limitations on the total electron dose and exposure time. Inelastically-scattered electrons do not contribute to image contrast; they can be removed by an energy filter or otherwise they contribute to shot noise. The amount of inelastic scattering with ice increases with the thickness of ice, and it is therefore helpful to have a thin ice layer. 

Elastic scattering is the physical phenomenon that gives rise to tomographic projections. For cryo-EM imaging of biological molecules the most important imaging mode
is phase contrast. The phase of an electron wave is shifted by its passage near the nucleus of each atom and it is elastically scattered (its energy is not changed). The positive potential it experiences (a few tens of volts over a path length on the order of an angstrom) results in a small phase advance on the order of a milliradian. The average inner potential of water (or vitreous ice) is about 5 V; for lipid
bilayers, about 6.5 V; for protein, about 7.5 V. It is the contrast between protein and ice that we are usually imaging. ``Phase-contrast'' and ``interference-contrast'' devices are used in light microscopy to see cultured cells (Zernike, Nobel Prize 1953). The primitive way to get contrast by an electron microscope is to defocus the objective lens. The process by which this phase-contrast image is formed is a bit complicated. One way do derive it is by use of the Fresnel propagator, the integral that shows how wavefronts develop in free space. The Fourier transform of the
Fresnel propagator yields the contrast transfer function (CTF) directly. Another way to derive it is based on diffracted electron waves, which we use here.

\subsection{Phase contrast imaging and the contrast transfer function}
\label{sec:phase}

In elastic scattering, electrons are scattered by the molecule in an angle that depends on the spatial frequency $k$ of the electrostatic potential the molecule creates. Consider a plane at a distance $z$ from the specimen, which will be the plane where the objective lens is focused. Negative values of $z$ refer to planes below the specimen, that is between the specimen and the objective lens. Because waves scattered at different angles traverse different distances between the specimen and the plane, they undergo different phase shifts. Classical diffraction theory implies that to first approximation this phase shift equals $\pi z \lambda |k|^2$ where $|k|$ is the radial frequency. In addition,
the weak phase approximation models the phase shift as the creation of a diffracted wave with
a phase shift (with respect to the unscattered wave) of $\frac{\pi}{2}$. The overall phase shift is $\Delta \varphi = \frac{\pi}{2} + \pi z \lambda |k|^2$.
The total wave function $\psi$ is the interference of the scattered wave $\psi_s$ with the unscattered wave $\psi_u$, with their superposition being $\psi = \psi_s + \psi_u$. The probability of an electron being detected by the detector is proportional to $|\psi|^2$. Suppose $\epsilon(k)$ is the frequency content of the sample, satisfying $\epsilon \ll 1$. Then,
\begin{equation}
|\psi|^2 = |\psi_u + \psi_s|^2 = |\psi_u|^2 |1 + \epsilon(k) e^{\imath \Delta \varphi}|^2 \approx |\psi_u|^2 (1 + 2\epsilon(k)\cos(\Delta \varphi) )
\end{equation}
The CTF $H$ is the deviation from 1, given by
\begin{equation}
\label{eq:simplectf}
H(k) = \cos(\Delta \varphi) = \sin (-\pi z \lambda |k|^2)
\end{equation}
Notice that the contrast is vanishing for an in-focus microscope ($z=0$). The point spread function (PSF) $h$ is the inverse Fourier transform of the CTF.

Let $f$ by the electrostatic potential of the sample. Ignoring noise and other modeling effects that are later discussed, the micrograph $I$ is given by
\begin{equation}
\label{eq:conv}
I(x,y) = (h \ast P f) (x,y)
\end{equation}
where $P$ is the tomographic projection operator defined by
\begin{equation}
\label{eq:tomo}
P f = \int_{-\infty}^\infty f(x,y,z) \,dz,
\end{equation}
where $z$ is the direction of the beaming electrons and $\ast$ denotes convolution: $(h \ast g) (x,y) = \int h(x-u, y-v) g(u,v)\,du\,dv$. The Fourier transform takes a convolution into a regular product, thus
\begin{equation}
\label{eq:conv-fourier}
(\mathcal{F}I)(k_x,k_y) = H \cdot \mathcal{F}Pf,
\end{equation}
where $\mathcal{F}$ denotes the Fourier transform.

The CTF oscillates between positive and negative values: some frequency bands get positive contrast, while others are negated. The CTF starts negative, therefore particle images in micrographs look dark rather than bright. Crucially, the CTF vanishes at certain frequencies (known as ``zero crossings") and information in the original projection images is completely lost at such frequencies. As a result, images at multiple defocus levels are needed for either 2D or 3D image recovery. Notice that CTF oscillations are more frequent at high frequencies as the phase grows as  $|k|^2$.
Larger defocus gives more contrast at low frequencies, but high frequency features are affected by the more frequent oscillations.

The power spectrum of the image is given by the squared magnitudes
\begin{equation}
|(\mathcal{F}I)|^2(k_x,k_y) = H^2 \cdot |\mathcal{F}Pf|^2
\end{equation}
(notice that $H$ is real-valued and it is therefore not required to take absolute value before squaring.)
It follows that the power spectrum vanishes on rings of radial frequencies that correspond to zero crossings of the CTF. The pattern of dark and bright rings in the power spectrum image is known as Thon rings, and facilitates the identification of the CTF (see (\textbf{Figure \ref{fig1}C})

We have been able to derive the CTF considering the electron wavefunction $\psi$ at a plane near the specimen; this is possible because the microscope's objective lens has the effect of reproducing the magnitude and phase of $\psi(x,y)$ at the conjugate image plane. Assuming an ideal lens and magnifaction factor $M$ the image wavefunction is,
\begin{equation}
\psi_M(x,y)=\frac{1}{M^2}\psi(\frac{-x}{M},\frac{-y}{M})
\end{equation}
and unscattered and scattered electron waves are also reproduced, with appropriate scaling, at the image plane. Hence, taking magnification into account, the CTF (eqn. \ref{eq:simplectf}) holds at the image plane.

Aberrations in the objective lens modify the CTF of the image. Taking astigmatism into account, the defocus $z$ is modeled by the three parameters $z_1, z_2, \alpha_z$ as
$z = \frac{1}{2} (z_1 + z_2) + \frac{1}{2}(z_1 - z_2) \cos (2 (\alpha_k - \alpha_z))$, where $\frac{1}{2}(z_1+z_2)$ is the average defocus, $\frac{1}{2} (z_1-z_2)$ is the astigmatism of the
objective lens, and $\alpha_z$ is the astigmatism angle (here, $\alpha_k$ is the angle of the frequency $k$ in polar coordinates). The astigmatism is typically small so that the CTF is approximately circularly symmetric.

A more accurate model for the CTF includes a $|k|^4$ term for a lens aberration in which scattered waves at high angles are overly focused by the objective lens. By analogy with optical lenses this is called spherical aberration, with coefficient $C_s$. Another term models the small fraction $w$ of amplitude contrast (beam attenuation) that arises from the loss of electrons that are elastically scattered at high angles and lost.  The commonly-used model for the CTF, valid up to a resolution of 2.5 $\rm{\AA}$ or so \cite{erickson1971measurement,wade1992brief} includes these terms:
\begin{equation}
H(k) = \sin(-\pi z \lambda |k|^2 + \frac{\pi}{2} C_s \lambda^3 |k|^4 - w) E(|k|),
\label{eq:ctf}
\end{equation}
where $E(|k|)$ is an envelope function that models the blurring due to imperfections in the imaging process. Of the various terms the user has substantial control only of the defocus $z$, which is invariably chosen to be positive (the so-called underfocus condition) to match the sign of $w$ and to mitigate the spherical aberration term.

The envelope function is often assumed to be Gaussian of the form $E(|k|) = e^{-B |k|^2 / 4}$, where $B$ is the so-called ``$B$-factor'' or ``envelope-factor''. $B$ has units of $\textsf{nm}^2$ or $\rm{\AA}^2$. The envelope function has the effect of blurring the image. Good cryo-EM images have $B$ values of $60-100~\rm{\AA}^2$, but even these values are not so good.
At $60~\rm{\AA}^2$ spatial frequencies of $4~\rm{\AA}$ are attenuated to $1/e$ of their original
amplitude, and the power in the signal (the square of the amplitude) is reduced
to about $1/10$ of the original value. Higher spatial frequencies are attenuated
even more.

The envelope function is the result of several factors. A typical defocus of $1~\mu \textsf{m}$ is about 500,000 wavelengths away from the specimen.
Now suppose that the effective electron source size is such that some of the
incident electrons follow a slightly different path than others. A typical situation in
a microscope with a tungsten filament source would be that the incident
electrons follow paths that differ in angle by $10^{-3}$ radians.
These different paths can blur out the image of high-resolution features at large
distances from the specimen. The variation in electron path is called spatial
incoherence.
For example, suppose the specimen has a periodicity $d=1~\textsf{nm}$. At a defocus
of $1~\mu \textsf{m}$ this results in differences in intensity with this same periodicity.
But the periodic pattern imaged by electrons at $10^{-3}$ radians will be shifted by
$10^{-3} \times 1~\mu \textsf{m} = 1~\textsf{nm}$ compared to the pattern imaged with zero angle (traveling
along the $z$-axis).
Thus if the paths of the incident electrons have random angles in this range, the
$1~\textsf{nm}$ pattern will be completely washed out.
This is why the field-emission electron gun is so important: it allows the effective
electron source size to be so small that angular spreads of $10^{-5}$ or $10^{-6}$ radians are attainable, which in turn allows high resolution at high defocus values.

Two other major contributions to the envelope function are beam-induced specimen movement (considered in the next section) and radiation damage. Although radiation damage is the random breaking of bonds and general incineration of the macromolecules in the specimen, it is routinely modeled simply as a decay of high-frequency image information with increasing electron dose.

\subsection{Direct electron detectors and beam induced motion}

Images are acquired by a detector. In the old days, experimentalists would use photographic film, which would then be scanned for digitization. Film was mostly replaced in the early 2000s by CCD cameras, but the resolution obtained using these detectors was almost always worse than $6 \rm{\AA}$, unless using very large datasets for highly symmetric molecules like a virus for which a $3 \rm{\AA}$ resolution was the state-of-the-art in 2012. This was rather disappointing, because theoretical studies predicted that cryo-EM should achieve much better resolution with significantly fewer particles. Cryo-EM was often referred to by the unflattering term ``blob-ology'' as it could only produce low-resolution structures that demonstrated the overall shape without important finer structural detail \cite{smith2014beyond}.

The situation changed dramatically around 2012 with the introduction of direct electron detector (DED) cameras \cite{li2013electron,MCMULLAN20161}. Unlike CCD cameras, where electrons are first converted to photons and photons are converted again to an electric signal, DED cameras detect electrons directly and can even detect a single electron hit. This gives them higher detective quantum efficiency (DQE) and therefore better SNR compared with CCD cameras. Moreover, DED cameras have a high frame rate and can operate in a video mode, for example, recording 40 frames during 2 seconds of exposure time.

Recording a video of static molecules that are maintained at liquid nitrogen temperature may seem pointless: If particles do not move during the imaging process, then individual frames can simply be averaged to produce an equivalent still image. Surprisingly, early experiments using DED cameras showed that there is movement of the specimen between individual movie frames, with particles exhibiting movements as large as $20 \rm{\AA}$ \cite{brilot2012beam}. The largest displacements are often associated with the first few frames of the movie. This movement of particles due to exposure is referred to as ``beam induced motion''. Although there is no consensus on the physical origin of the motion and its modeling, one plausible explanation is that the motion is the result of energy transfer from the electron beam to the specimen: the molecules are irradiated by inelastic scattering which in turn may induce mechanical pressure on the supporting ice layer that wiggles and buckles as a result.

The image blur associated with beam induced motion can be corrected by alignment of individual movie frames, therefore significantly enhancing image quality and the overall resolution. Motion-corrected images enjoy a smaller B-factor, Thon rings are visible at higher radial frequencies (indicating better resolution), and the contrast is better. Consequently, the introduction of DED cameras has entered the cryo-EM field into a new era, because it has enabled structure determination at significantly better resolutions and from smaller macromolecules. Resolutions of $3-4 \rm{\AA}$ were first reported \cite{cheng-trpv1,amunts2014structure} and  structures have recently been determined to about $1.5 \rm{\AA}$ resolution \cite{bartesaghi2018atomic}. This transformative change in the power of cryo-EM is known as the ``Resolution Revolution'' \cite{kuhlbrandt}.

\section{PROCESSING MICROGRAPHS}

The computational pipeline in single-particle cryo-EM can be roughly split into two stages. The first stage consists of methods that operate at the level of entire micrographs such as motion correction, CTF estimation, and particle picking. The second stage involves methods that operate on individual particle images such as 2D classification, 3D reconstruction, and 3D heterogeneity analysis. The two stages are also distinguished by the type of approaches. Methods in the first stage mainly rely on classical image and signal processing, whereas those in the second stage are mostly based on statistical estimation and optimization. In this section we focus on micrograph processing.

\subsection{Motion correction}
Motion correction refers to translational alignment and averaging of movie frames.
Early methods for motion correction find the relative motion between any pair of individual frames and then estimate the frame shifts by solving a least squares problem. However, the movement is not homogeneous across the field of view, with particles in different regions moving in different velocities (in terms of both direction and speed). The local motion can be estimated by cross-correlating smaller image patches. Assuming that the velocity field is smooth with respect to both location and time, the initial estimates are further fitted and interpolated using local low-degree  polynomials. After estimation of the velocity field, the frames are aligned and averaged \cite{ripstein2016chapter,zheng2017motioncor2}. It is also possible to compute dose weighted averaging, where the weights are derived from a radiation damage analysis \cite{grant2015measuring}.

\subsection{CTF estimation and correction}
Although the CTF may vary across the micrograph due to a slight tilt or due to particles at different heights within the ice layer, in most cases those possible variations are ignored to first approximation. The three defocus parameters of the CTF are then estimated from the power spectrum of the whole micrograph, utilizing the following relationship between the power spectral density (PSD) of the micrograph $PSD_I$ with the CTF $H$, the average PSD of the particles, carbon support and other contaminants $PSD_f$, and the PSD of the background noise $PSD_b$, given by
\begin{equation}
\label{eq:PSD}
PSD_I = H^2 \cdot PSD_f + PSD_b.
\end{equation}
In principle, estimating the parameters of $H$ from Eq.(\ref{eq:PSD}) is an ill-posed problem, because both $PSD_f$ and $PSD_b$ are unknown. In practice, the estimation is tractable due to the fact that $PSD_f$ and $PSD_b$ are slowly varying functions (and usually monotonic decreasing in the radial frequency), while $H^2$ is a rapidly oscillating function.

The CTF $H$ is approximately circularly symmetric, because astigmatism is typically small. Circularly averaging the estimated $PSD_I$ produces a cleaner radial profile which is a superposition of the slowly varying background term and the rapidly oscillating CTF-dependent term. The background term can be estimated by local averaging of the radial profile, by fitting a low-degree polynomial, or by linear programming methods. The local minima of the background subtracted PSD correspond to the zero crossings of the CTF from which the defocus value can be estimated. For noisier and more challenging micrographs for which finding local minima is too sensitive, the defocus can be determined using cross-correlation of the background subtracted PSD with the CTF (see \textbf{Figure \ref{fig1}D}). The two other defocus parameters related to astigmatism are determined in a similar fashion by cross-correlating the CTF with the background subtracted PSD in two dimensions \cite{rohou2015ctffind4,zhang2016gctf}.

After estimating the CTF, it is sometimes desirable to perform CTF correction of the images, that is, to undo the effect of the CTF on the images. This is useful, for example, for obtaining a better visual sense of how projection images really look like without CTF distortion. The CTF distorts the images in several ways, including delocalizing images by increasing their support size.

CTF correction requires deconvolution by the PSF $h$ (see Eq.(\ref{eq:conv})), or equivalently, in the Fourier domain, inversion by the CTF $H$ (see Eq.(\ref{eq:conv-fourier})). This is an ill-posed problem due to the zero crossings of the CTF and the decay of the envelope function at high frequencies. A popular approach for CTF correction is ``phase flipping'', by which the filter $\operatorname{sign}(H)$ (which takes the values $1$ and $-1$, depending on the sign of $H$) is applied to the images. Phase flipping corrects for the phase of $H$, but not for its varying amplitude.
It is impossible to fully correct for the CTF of an individual micrograph, because there is information missing at certain frequencies where the CTF is zero; and near the zeroes and at high frequencies, where there
is information, it is nevertheless often unusable because its amplitude is so low. Beyond its simplicity, another advantage of phase flipping is that it preserves the noise statistics.

Another frequently used method for CTF correction is Wiener filtering \cite{sindelar2011adaptation}, which also attempts to correct for the amplitude of $H$. The Wiener filter minimizes the mean squared error among all linear filters and employs the spectral signal-to-noise ratio (SSNR), which is the function of the radial frequency given by $\text{SSNR} = \frac{PSD_f}{PSD_b}$ (notice that $PSD_f$ and $PSD_b$ are byproducts of the CTF estimation procedure). Compared with ``phase flipping'', amplitude correction is better accounted for in Wiener filtering, but information missing at zero crossings cannot be recovered.

In more advanced stages of the computational pipeline, such as 2D classification and 3D reconstruction, the problem of information loss at the zero crossings is overcome by combining data taken at various defocus values.  This way the zeros from one image are filled in by data from others. Wiener filtering can be used to combine information from multiple defoci, as well as invert the effects of the CTF. Such information aggregation happens when 2D class averages are formed following the step of 2D classification, and when 3D reconstruction is performed from images with varying CTFs.

\subsection{Particle picking}
\label{sec:picking}

Current 3D reconstruction procedures require as input individual particle images rather than entire micrographs. Therefore, particle images need to be selected from the micrographs while discarding regions that contain noise or contaminants. ``Particle picking'' is the computational step of segmenting and boxing out particle images from micrographs. The reliability of particle detection depends on the particle size: the smaller the molecule, the lower the SNR and the image contrast. Therefore, particles of samples with small molecular mass cannot be reliably detected and oriented due to poor SNR \cite{sigworth2004classical}. The detection threshold was recognized early on as a crucial limiting factor in SPR \cite{henderson1995potential,glaeser1999electron}.
To date, the biological macromolecules whose structures have been determined using cryo-EM usually have molecular weights greater than 100 kDa. Cryo-EM using defocus contrast cannot be applied to map molecules with molecular mass below a certain threshold ($\sim 40$ kDa at present), because particles of samples with small molecular mass cannot be reliably detected and oriented.

Image contrast depends crucially on the CTF. Large defocus increases low-frequency content and improves image contrast, making detection of particles easier. On the other hand, high-frequency features are mitigated at large defocus, and it is therefore required to use images taken at low defocus for which particle picking is more challenging.

Image contrast can be significantly improved using a Volta phase plate (VPP) that effectively shifts the phase of the CTF, thus boosting information content at low frequencies \cite{danev2016cryo,khoshouei2017cryo}. However, VPP are often unstable and suffer from phase drift. There is an ongoing debate regarding the added value of VPP, and other physical mechanisms for phase shift are being actively explored.

Hundreds of thousands of particle images are often required for heterogeneity analysis and high resolution reconstruction. Manual selection of such a large number of particles is tedious and time consuming. Semi-manual and automatic approaches for particle picking are therefore becoming more common.

Many procedures for particle picking are based on template matching \cite{frank1983automatic,chen2007signature,scheres2015semi}, where a small number of template images are cross-correlated with all possible micrograph windows, and windows with high correlation values are selected. The templates can be predefined, such as Gaussian blobs or their difference \cite{voss2009dog}, or they can be determined by other means. For instance, low resolution (e.g., at $\sim 40 \rm{\AA}$) projection images obtained by negative staining can be used as templates. Another alternative is to manually select a relatively small number of particles, e.g., 1000 or 10000 particles, and compute from those a smaller number of 2D class averages (say, between 10 to 100) to be used as templates.

Although template matching is simple and fast, it has the potential risk of model bias in the sense that the particular choice of templates dictates which windows are selected as particles and may bias the final reconstruction towards having projections similar to the templates. In cryo-EM folklore this risk is referred to as ``Einstein from noise'' \cite{shatsky2009method,henderson2013avoiding}: when an image of Einstein is used as template to select particles from pure noise micrographs, the average of selected particles mimics the original image of Einstein. To address this issue, an automated template-free procedure was recently proposed \cite{heimowitz2018apple}.

Following the success of deep neural networks in image recognition and natural language processing, among others, their application to particle picking has also been considered recently \cite{wang2016deeppicker,wagner2019sphire,bepler2019positive}. The hope is that a deep net architecture can be trained with a small number of samples to match the quality of manual selection but be much faster. Deep nets are also prone to model bias. This is an active area of research with room for improvement in terms of automation, speed, and quality of detection. Eliminating the step of particle picking altogether by reconstructing directly from the micrographs is another exciting research avenue.

\begin{figure}
\includegraphics[width=1.0\textwidth]{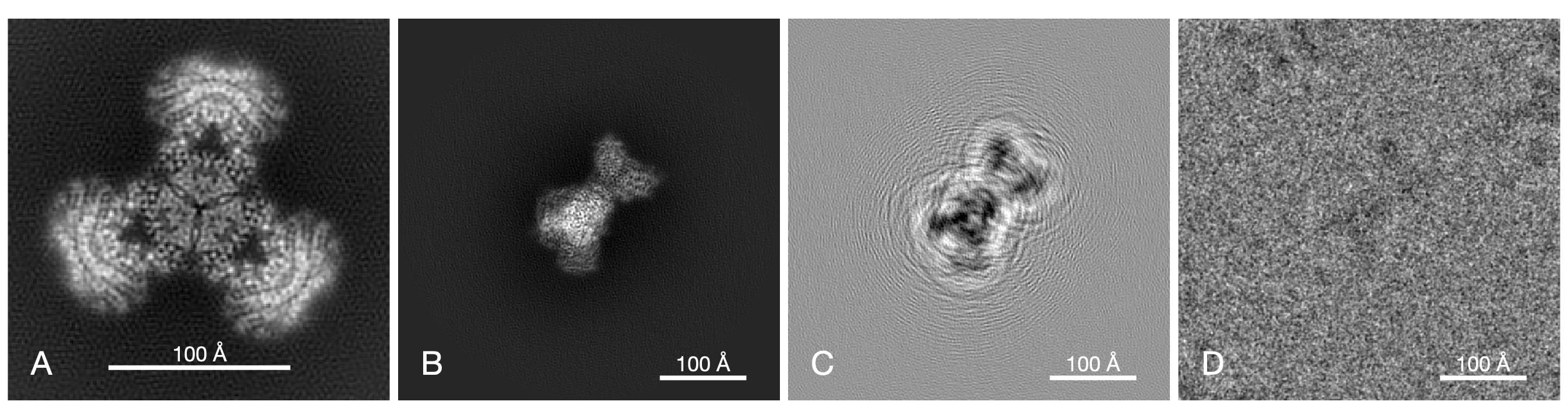}
\caption{Model of image formation by defocus contrast. {\bf A}, a projection of the experimental density map of PaaZ. {\bf B}, projection of the map at the best-match orientation determined from a particle image. {\bf C}, the same projection, but filtered by the CTF. The substantial defocus ($2.6 \mu$m) results in considerable delocalization of information, resulting in fringes surrounding the particle. {\bf D}, the corresponding experimental image of a particle. The large size of the particle image ($440 \times 440 \rm{\AA}$) is necessary to include the delocalized information up to a spatial frequency of about $0.4 \rm{\AA}^{-1}$.}
\label{fig2}
\end{figure}

\section{THREE-DIMENSIONAL RECONSTRUCTION}

Three-dimensional reconstruction is the focal point of the computational pipeline of SPR, as in this step the three-dimensional electrostatic potential of the molecule needs to be estimated from the picked particle images. We therefore devote to this computational step the most attention. This section provides the mathematical, statistical, and computational aspects of 3D reconstruction.

Particle images are noisy two-dimensional tomographic projection images of the molecules taken at unknown viewing angles. Moreover, the projections are slightly off-centered, distorted by the CTF, and may be affected by other sources of variability such as variations in the ice thickness, all of which further complicate the reconstruction. Another difficulty is that the stack of images often contains a significant number of outliers that were falsely identified as particles. Such non-particle images might be images with more than one particle, images with only a part of a particle, or images of pure noise.

Related to 3D reconstruction in SPR is the reconstruction problem in classical computerized tomography (CT) that arises in medical imaging. In both cases the 3D structure needs to be determined from tomographic projection images. There are, however, some key differences between the two problems. First, the pose parameters (i.e., the viewing angles and centers) of projection images are known in CT, but they are unknown in SPR. CT reconstruction from projection images with known pose parameters is a linear inverse problem for which there exist mature solvers \cite{herman2009fundamentals}. The missing pose parameters in SPR makes it a much more challenging nonlinear inverse problem. Second, the level of noise in SPR is significantly higher compared with CT. As a result, a large number of images is required in order to average out the noise. This lends itself to computational challenges that are associated with processing large datasets.

\subsection{Maximum likelihood estimation}

At its core, 3D reconstruction is a statistical estimation problem. The electrostatic potential $\phi$ of the molecule needs to be estimated from $n$ noisy, CTF-affected projection images $I_1,I_2,\ldots,I_n$ with unknown viewing angles, modeled as
\begin{equation}
\label{eq:basic-model}
I_i = h_i \ast P R_i \circ \phi + \epsilon_i, \quad i=1,\ldots,n.
\end{equation}
Here, $h_i$ is the PSF of the $i$'th image (the inverse Fourier transform of the CTF $H_i$), $P$ is the tomographic projection operator (defined in Eq. (\ref{eq:tomo})), $R_i$ is an unknown $3\times 3$ rotation matrix (i.e., an element of the rotation group $SO(3)$) representing the unknown orientation of the $i$'th particle, and $R\circ \phi$ denotes rotation of $\phi$ by $R$, that is, $(R \circ \phi)(\mb{r}) = \phi(R^{-1} \mb{r})$ for $\mb{r}=(x,y,z)$. The noise is represented by $\epsilon_i$ and assumptions on the noise statistics are discussed later. Other effects, such as varying image contrast, varying noise power spectrum, uncertainties in CTF (e.g., defocus value), and non-perfect centering can be included in the forward model (\ref{eq:basic-model}) as well, but we leave them out for simplicity of exposition. \textbf{Figure \ref{fig2}} is an illustration of the image-formation model.

From a structural biologist's standpoint, the 3D structure $\phi$ is the only parameter of interest; the rotations $R_1,\ldots, R_n$ are nuisance parameters. While the actual values of rotations definitely impact the quality of the reconstruction, they do not carry any added value about the molecular structure. In fact, all other model parameters besides $\phi$ such as image shifts, per-image defocus value, noise power spectrum, and image normalizations are considered nuisance parameters.

Maximum likelihood estimation is a cornerstone of statistical estimation theory as it enjoys several appealing theoretical properties \cite{wasserman2013all}. Under certain mild technical conditions the maximum likelihood estimator (MLE) is a consistent estimator and it is asymptotically efficient in the sense that it achieves the Cram\'er-Rao bound (in other words, it has the smallest possible variance in the limit of large sample size). At first glance it seems as if there are two alternative ways of defining the MLE. One way would be to find $\phi$ and $R_1,\ldots,R_n$ that maximize the likelihood function. The other way would be to find $\phi$ that maximizes the marginalized likelihood function, where marginalization is over all nuisance parameters (such as $R_1,\ldots,R_n$). The latter is the ``correct'' way in the sense that the technical conditions that guarantee consistency and efficiency of the MLE are satisfied. Indeed, the number of parameters must be fixed and independent of the size of the data to ensure consistency and efficiency. But if rotations are also estimated, then the number of model parameters grows indefinitely with the number of images. The failure of MLE in this case is sometimes regarded as overfitting due to the large number of model parameters. A famous example of the inconsistency of the MLE in the case of increasingly many parameters is the Neyman-Scott ``paradox'' \cite{neyman1948consistent}.

Assuming white Gaussian noise with pixel variance $\sigma^2$ (again, this restrictive assumption is only made for exposition purposes), the likelihood function is
\begin{equation}
L_n(\phi,R_1,\ldots,R_n) = \Pr\{I_1,\ldots,I_n \,|\, \phi,R_1,\ldots,R_n \} = \prod_{i=1}^n \frac{1}{(2\pi \sigma^2)^{p/2}} e^{-\frac{\|I_i - h_i\ast P R_i \circ \phi \|^2}{2\sigma^2} },
\end{equation}
where $p$ is the number of pixels in an image. The marginalized likelihood function $L_n(\phi)$ is obtained by integrating $L_n(\phi,R_1,\ldots,R_n)$ over the nuisance parameters as
\begin{eqnarray}
L_n(\phi) &=& \int_{SO(3)^n} L_n(\phi,R_1,\ldots,R_n) dR_1 dR_2 \cdots dR_n \\
&=& \int_{SO(3)^n} \left( \prod_{i=1}^n  \frac{1}{(2\pi \sigma^2)^{p/2}} e^{-\frac{\|I_i - h_i\ast P R_i \circ \phi \|^2}{2\sigma^2} }\right) dR_1 dR_2 \cdots dR_n \\
&=& \prod_{i=1}^n \int_{SO(3)}  \frac{1}{(2\pi \sigma^2)^{p/2}} e^{-\frac{\|I_i - h_i\ast P R \circ \phi \|^2}{2\sigma^2} } dR.
\end{eqnarray}
The marginalized log-likelihood is given by
\begin{equation}
\ell_n(\phi) = \log L_n(\phi)  = - \frac{np}{2}\log (2\pi \sigma^2) + \sum_{i=1}^n \log \int_{SO(3)}   e^{-\frac{\|I_i - h_i\ast P R \circ \phi \|^2}{2\sigma^2} } dR.
\end{equation}
Since the first term is independent of $\phi$, the maximum marginalized likelihood estimator $\hat{\phi}^{\text{MLE}}$ is defined as
\begin{equation}
\hat{\phi}^{\text{MLE}} =  \argmax_\phi \sum_{i=1}^n \log \int_{SO(3)}   e^{-\frac{\|I_i - h_i\ast P R \circ \phi \|^2}{2\sigma^2} } dR.
\end{equation}

Finding $\hat{\phi}^{\text{MLE}}$ therefore requires solving a high-dimensional, nonlinear, and non-convex optimization problem. For example, for a 3D structure represented by $100 \times 100 \times 100$ grid points, the optimization is over a space of $10^6$ parameters. For such high-dimensional optimization problems, classical second-order algorithms such as Newton's method may be ill suited, because executing even a single iteration may be too costly. This leaves us with first-order methods (where only the gradient is computed or approximated) and the expectation-maximization
(EM) algorithm.  To avoid any acronym confusion, we use ML-EM to refer to maximum-likelihood expectation-maximization. Due to its iterative nature, ML-EM is also widely referred to as iterative refinement. Both ML-EM and first-order methods are popular workhorses for 3D reconstruction.

\subsection{Expectation-Maximization}
ML-EM is tailored to missing data problems such as the cryo-EM reconstruction problem in which the rotations are latent variables \cite{dempster-laird,sigworth1998maximum}. ML-EM starts with some initial guess $\hat{\phi}^{(0)}$ and then iterates between two steps until convergence: the expectation (E) step, and the maximization (M) step. In the E-step of iteration $t$, a conditional distribution of the latent variables is computed given the current estimate of the parameter of interest $\hat{\phi}^{(t)}$. That is, for each latent variable $R_i$, the parameter space $SO(3)$ is discretized with $S$ fixed rotations $U_1,\ldots,U_S \in SO(3)$, and a vector of responsibilities $\gamma_i^{(t)}$ (also known as weights or probabilities) of length $S$ is computed as
\begin{equation}
\gamma_i^{(t)} (s) = \frac{e^{-\frac{\|I_i - h_i\ast P U_s \circ \hat{\phi}^{(t)} \|^2}{2\sigma^2} }} {\sum_{s'=1}^S e^{-\frac{\|I_i - h_i\ast P U_{s'} \circ \hat{\phi}^{(t)} \|^2}{2\sigma^2} }}, \quad s=1,\ldots,S.
\end{equation}
Note that the denominator is simply a normalization to ensure the responsibilities sum to one.
In the $M$-step, a new estimate $\hat{\phi}^{(t+1)}$ is defined as the maximizer of the log-likelihood of the images with the given responsibilities
\begin{equation}
\label{eq:least-squares}
\hat{\phi}^{(t+1)} = \argmin_{\phi} \sum_{i=1}^n \sum_{s=1}^S \gamma_i^{(t)}(s) \|I_i - h_i\ast P U_s \circ \phi \|^2.
\end{equation}
This is a linear least squares problem for which the solution $\hat{\phi}^{(t+1)}$ can be obtained by any of the existing methods for CT reconstruction, such as the Fourier gridding method \cite{penczek2004gridding}, non-uniform fast Fourier transform (NUFFT) methods \cite{nufft,wang2013fourier}, or algebraic reconstruction techniques (ART) \cite{marabini19983d}.

ML-EM is guaranteed to increase (more precisely, not decrease) the marginal likelihood in each iteration, and to converge to a critical point. However, it does not necessarily converge to the global optimum. In practice, either one needs to have a good initial guess, or run ML-EM multiple times with different initializations, and pick the solution
with the largest likelihood value.

\subsubsection{Computational complexity of ML-EM}

The E-step requires the computation of the responsibility vectors $\gamma_i^{(t)}$.
For images of size $L \times L$, the latent variable space of $SO(3)$ is discretized with $S = O(L^3)$ rotations. The discretization is taken as $O(L^2)$ different viewing directions on the unit sphere, each of which is in-plane rotated $O(L)$ times.

To compute tomographic projections of the current estimate $\hat{\phi}^{(t)}$ one employs the Fourier projection slice theorem \cite{natterer}: For sampling $O(L^2)$ slices of size $L\times L$, the Fourier transform of $\hat{\phi}^{(t)}$ needs to be evaluated on $O(L^4)$ off-grid points. This is computed efficiently using NUFFT in $O(L^4)$ \cite{barnett2019parallel}.

The cost of projection-matching a single noisy image $I_i$ with a single projection image $h_i \ast PU_s \circ \hat{\phi}^{(t)}$ and all its in-plane rotations is $O(L^2 \log L)$. This is achieved by resampling the images on a polar grid (e.g., using NUFFT) and computing the correlations over concentric rings by employing the convolution theorem and using the standard 1-D FFT. Thus, the cost of computing all responsibility vectors for $n$ images is $O(n L^4 \log L)$. This step clearly dominates the cost of generating the projection images.

The M-step involves a weighted tomographic reconstruction from the noisy images. The solution to the least squares problem Eq.(\ref{eq:least-squares}) can be formally written as
\begin{equation}
\hat{\phi}^{(t+1)} = \left(\sum_{i=1}^{n} \sum_{s=1}^{S} \gamma_i^{(t)}(s) (PU_s)^T h_i \ast h_i\ast P U_s   \right)^{-1} \left( \sum_{i=1}^n \sum_{s=1}^S \gamma_i^{(t)}(s) U_s \circ P^T h_i \ast I_i \right),
\end{equation}
where $P^T$ is the back-projection operator. In words, the least squares solution involves back-projecting all images in all possible directions followed by a matrix
> inversion. For computing the back-projection of an image for all possible rotations, it is more efficient to fix a viewing direction and first compute a weighted average (with respect to the latent probabilities) of the image and its in-plane rotations in $O(L^2 \log L)$. The cost for repeating for all images and all viewing directions is $O(n L^4 \log L)$. The matrix to be inverted can be shown to be a convolution kernel and its computation also takes $O(n L^4 \log L)$ using NUFFT \cite{wang2013fourier}. The linear system can be solved iteratively using conjugate gradient (CG). The convolution kernel has a Toeplitz structure as a matrix and it can therefore be applied fast using FFT. Each iteration of CG costs $O(L^3 \log L)$, and the overall cost of CG iterations is negligible compared to that of previous steps.

We conclude that the cost of projection-matching (E-step) is comparable to the cost of tomographic reconstruction (M-step), both given by $O(n L^4 \log L)$.
Typical sizes of $n=10^5$ and $L=200$ give $nL^4 = 1.6 \times 10^{14}$. Keeping in mind that images are off-centered, another factor of 100 can easily be included to account for searching in a $10\times 10$ square of possible translations, and a factor of at least $10$ for the logarithmic factors and constants of the NUFFT. The number of flops can easily top $10^{17}$. The main factor responsible for the ballooning computational cost is the grid size $L$, or equivalently the resolution, due to the $L^4$ scaling.

\subsubsection{Accelerated ML-EM}

Given the high computational cost of ML-EM it is no surprise that considerable effort was invested in modifying the basic approach in order to reduce the overall computational cost. The most common way to cut the cost of the tomographic reconstruction step is by sparsification of the responsibility vectors (i.e., by setting small weights to zero). This thresholding is justified whenever the conditional distribution of the rotations is peaky around its mode \cite{frealign}. That is, whenever there is a high confidence in rotation assignment. In such cases, hard assignment of rotations (i.e., assigning the entire weight to the best-match rotation, and zero weight to all other rotations) is much cheaper compared to soft assignment and still provides a similar output. It should be noted that hard assignment of rotations was in fact proposed and applied for SPR long before ML-EM \cite{harauz1983direct,harauz1984nucleosome}. Another way to cut down the cost of the M-step is by using the gridding technique or a crude interpolation instead of the more accurate NUFFT. Given the noise in the data, the approximation error of the gridding method is often tolerable.

Computational savings of both the E and M steps are obtained by limiting the search space of the hidden variables. For example, the previous ML-EM iteration provides a good idea of the latent variable probabilities and one can set a lower bound to cut the search space. This is the basis of the branch and bound method \cite{brubaker}.

The latent variable space can also be discretized at different levels of granularity. In adaptive EM \cite{tagare2010adaptive,scheres-relion} several grids are used for latent variable space of rotations and shifts. After a first swap over the coarse grid, only grid points that account for $99.9\%$ (or some other fixed threshold) of the overall probability are kept, and then the search proceeds to the finer level that corresponds to these points.  Adaptive EM does not help so much in initial iterations when the latent distributions are flat, but can produce an order of magnitude acceleration near convergence when the probabilities become more ``spiky''.

Subspace-EM \cite{dvornek2015subspaceem} accelerates ML-EM by reducing the dimensionality of the images using principal component analysis (PCA). Correlations between the projection images and the eigen-images (principal components) are pre-computed. Then, correlations between projection images and the noisy experimental images are quickly evaluated as linear combinations of the pre-computed correlations.

Even with all these modifications and heuristics, iterative refinement is often a computationally expensive step, especially if a high resolution reconstruction or heterogeneity analysis are performed on large datasets \cite{scheres2009maximum,scheres2016processing}. It is also becoming more common for the reconstruction procedure to be applied to individual movie frames and to introduce more latent variables such as per-image defocus \cite{frealign,bartesaghi2018atomic} in order to reach higher resolution and squeeze out as much as possible out of the data \cite{zivanov2019bayesian}. To accelerate such large-scale computations, specialized GPU implementations have been recently developed \cite{brubaker,kimanius,zivanov2018new}.

\subsection{Bayesian Inference}

The problem of tomographic reconstruction (even with known pose parameters) can be ill-posed, due to poor coverage of slices in the Fourier space and small values of the CTF at certain frequencies. The MLE can therefore have a large mean squared error (MSE), even though it is consistent and asymptotically efficient. This is a well-known problem of the maximum likelihood framework which is not specific to SPR.

A classic solution to counter the bad conditioning is using regularization which can significantly reduce the MSE. In effect, regularization incorporates a prior on the parameters of interest and the prior information is weighted accordingly with the observed data.

Without regularization, the ML-EM procedure for reconstruction was empirically observed to often produce 3D structure with artificial content at high frequencies. This is a clear indication of over-fitting. A prior on the model parameters may help mitigate these problems. This is the basic idea behind Bayesian inference and it has proven to be quite powerful in SPR \cite{scheres2012bayesian,scheres-relion}.

An important concept in Bayesian inference is that of the posterior distribution. Suppose $\Pr(\phi)$ is a prior distribution on the parameter $\phi$. Using Bayes' law, the posterior probability of the parameters given the data $X$ is
\begin{equation}
\Pr(\phi|X) = \frac{\Pr(\phi,X)}{\Pr(X)} = \frac{\Pr(X|\phi) \Pr(\phi)}{\Pr(X)}
\end{equation}
The denominator $\Pr(X)$ is called the evidence, and is independent of $\phi$. Thus, maximizing the posterior probability is equivalent to maximizing $\Pr(X|\phi) \Pr(\phi)$.
Recall that $\Pr(X|\phi)$ is the likelihood. So, the maximum a-posteriori probability (MAP) estimate is obtained by maximizing the likelihood times the prior.

The posterior distribution $\Pr(\phi|X)$ can in fact be very powerful: If it can be somehow computed, then the mean, variance, and any statistic of interest of $\phi$ can be derived. This might be very useful for 3D reconstruction --- as we can obtain not just the mean volume, but also confidence intervals which can prove useful for model validation. The posterior can be used to define the minimum mean square error (MMSE) estimator, given by
\begin{equation}
\hat{\phi}^{\text{MMSE}} = \mathbb{E}[\phi|X] = \int \phi \Pr(\phi|X)\,d\phi,
\end{equation}
the derivation of which is analogous to the Wiener filter.

However, except for some specialized low-dimensional problems, calculating the posterior distribution is out of reach. There are clever techniques for sampling from the posterior distribution (e.g., Markov Chain Monte-Carlo methods), but they are currently not widely used in SPR. Variational inference is another interesting approach recently proposed for reconstruction and for approximating posterior distributions \cite{ullrich2019differentiable}.

The MAP estimate is obtained by maximizing the sum of the log-likelihood with the log-prior
\begin{equation}
\label{eq:MAP}
\hat{\phi}^{\text{MAP}} = \argmax_{\phi} \left(\log \Pr(X|\phi ) + \log \Pr(\phi)\right).
\end{equation}
MAP estimation is often regarded as the ``poor man's version of Bayseian inference'', since the only information obtained about the posterior distribution is its mode. The MAP estimate is a solution to an optimization problem. Sometimes it is the only practical choice for large-scale problems, for which it is too expensive to get more information about the posterior.

The expectation-maximization algorithm can also be applied to MAP estimation, and we refer to it as MAP-EM. It enjoys from similar theoretical guarantees, namely, the posterior probability increases in each iteration, and iterations converge to a critical point of the posterior. MAP-EM iterations consist of an E-step and an M-step. The E-step is the same as for ML-EM, and a soft projection-matching step is performed to compute conditional probabilities of the latent variables. In the M-step a new structure is reconstructed that maximizes the posterior (rather than the likelihood) given the conditional probabilities of the latent variables. Here is where the prior comes into play by regularizing the tomographic reconstruction step.

A popular prior in SPR is to assume \cite{scheres2012bayesian} that each frequency voxel of $\Phi$ (the Fourier transform of $\phi$) is an independent zero-mean complex Gaussian, with variance $\tau^2$:
\begin{equation}
\Pr(\Phi) = \prod_{l=1}^p \frac{1}{2\pi \tau_l^2 }\exp\left\{-\frac{|\Phi_l|^2}{2\tau_l^2} \right\}.
\end{equation}
The variance function $\tau^2$ is unknown and is treated as a model parameter, that is, it is refined during EM iterations via
\begin{equation}
\label{eq:tau}
(\hat{\tau}_l^2)^{(j+1)} = \frac{1}{2} |\hat{\Phi}_l^{(j+1)}|^2.
\end{equation}
In practice, $\tau$ is assumed to be a 1-D radial function, so its new value is obtained by averaging over concentric frequency shells.
The value of $\tau$ is actually multiplied by an ad-hoc constant $T$, recommended to be set to values between 2 and 4. The intuition is that the prior does not account for correlations between the Fourier coefficients, hence the diagonal part of the covariance matrix needs to be artificially inflated \cite{scheres2012bayesian}.

The Gaussian prior implies that the MAP 3D reconstruction update step effectively includes a Tikhonov regularization parameter inversely proportional to $\displaystyle{\frac{1}{\tau^2}}$ in the Fourier domain; equivalently, $\tau^2$ can be regarded as the SNR in the Wiener filtering framework. If the initial guess $\phi^{(0)}$ for the 3D structure is of low resolution, then Eq. (\ref{eq:tau}) suggests that $\hat{\tau}^{(0)}$ would be large only at low-frequencies, hence high frequencies would be severely regularized. As a result, the MAP-EM iterates would gradually increase the effective resolution of the 3D map. This behavior is often referred to as frequency marching, in which the resolution gradually increases as iterative refinement proceeds.

Masking is another popular prior in SPR. In masking, the support of molecule in real space is specified and values of $\phi$ outside the support are masked out and set to zero. Masking can push the resolution higher by effectively decreasing the parameter space and diminishing the noise. However, this comes at risk of introducing manual bias into the reconstruction. In principle, the mask should be provided before iterative refinement (for example, if the molecule is known to be confined to a geometrical shape that deviates considerably from a ball, e.g., a cigar-like shape). In practice, however, it is common to determine the mask after a first run of iterative refinement, and provide that mask to a second run of iterative refinement. This practice is an example of a mathematical ``inverse crime'' \cite{kaipio2006statistical} and is known to lead to overoptimistic results.

\section{AB-INITIO MODELING}

Iterative refinement methods for 3D reconstruction require an initial model $\phi^{(0)}$. Expectation-maximization is only guaranteed to converge to a critical point of the objective function. The non-convex landscape of the likelihood function (with or without regularization) implies that iterative refinement may not converge to the global maximum and convergence depends on the initial guess. It is therefore important to have a good initialization. Another benefit of a good initialization is reduction in running time: With a good initialization, fewer iterations are required to reach convergence, and the cost of each iteration is also cheaper because the distribution of the latent variables is no longer flat (therefore, the size of the search space can be reduced). In this section we review methods for estimating a low-resolution ab-initio model directly from the data.

\subsection{Stochastic gradient descent}
As mentioned earlier, computing the MLE or the MAP estimate requires large-scale optimization for which second-order methods are not suitable. Instead, expectation-maximization and first-order gradient methods are natural candidates for optimization. Classical gradient methods are likely however to get trapped in bad local optima. Moreover, the computation of the exact gradient is expensive as it requires processing all images at once. In stochastic gradient descent (SGD) only a single image or a few randomly selected images are used to compute an approximate gradient in each iteration \cite{brubaker}.

SGD exploits the fact that the objective function is decoupled into individual terms each of which depends on a single image. Indeed, the MAP estimate is the solution of the optimization problem
\begin{equation}
\hat{\phi}^{\text{MAP}} = \argmax_\phi \log \Pr(X | \phi) + \log \Pr(\phi) = \argmax_\phi \sum_{i=1}^n \left(\log \Pr(I_i | \phi) + \frac{1}{n} \log \Pr(\phi)\right).
\end{equation}
The log-posterior objective $f$ can be written as
\begin{equation}
f(\phi) = \sum_{i=1}^n f_i(\phi),\quad f_i(\phi) = \log \Pr(I_i | \phi) + \frac{1}{n} \log \Pr(\phi).
\end{equation}

Gradient based algorithms (often called ``batch'') are of the form $\phi^{(t+1)} = \phi^{(t)} + \alpha_t \nabla f(\phi^{(t)})$, where $\alpha_t$ is the step size (also known as the learning rate). For SGD, the gradient is approximated by a random subset (drawn uniformly) $\mathcal{I}_t$ (mini-batch) of $f_i$'s:
\begin{equation}
\phi^{(t+1)} = \phi^{(t)} + \alpha_t \frac{1}{|\mathcal{I}_t|}\sum_{i \in \mathcal{I}_t} \nabla f_i(\phi^{(t)})
\end{equation}
Only images in the batch are needed to compute the approximate gradient. There are several variants of SGD that can be more suitable in practice, such SAG and SAGA, among others \cite{bottou2018optimization}.

It is commonly believed that the inherent randomness of SGD helps to escape local minima in some non-convex problems. Note that due to its stochastic nature, SGD iterations never converge to the global optimum but rather fluctuate randomly. Hence, SGD is not useful for high-resolution 3D reconstruction. Instead, the main application of SGD is for low-resolution ab-initio modeling, and resolutions up to $10 \rm{\AA}$ have been successfully obtained using this method. Although strictly speaking SGD is not a fully ab-initio method because it also requires an initial guess $\phi^{(0)}$, it has been empirically observed to have little dependence on the initialization, and even random initializations perform well.

\subsection{Common-line approaches}

The Fourier slice theorem states that the 2D Fourier transform of a tomographic projection equals the restriction of the 3D Fourier transform of the molecule to a central slice perpendicular to the viewing direction \cite{natterer}. The Fourier slice theorem implies the common line property: the intersection of two (non-identical) central slices is a line. Therefore, for any pair of projection images, there is a pair of central lines (one in each image) on which their Fourier transforms agree. For non-symmetric generic molecular structures it is possible to uniquely identify the common-line, for example, by cross-correlating all possible central lines in one image with all possible central lines in the other image, and choosing the pair of lines with maximum cross-correlation. The common line pins down two out of the three Euler angles associated with the 3D relative rotation $R_i^{-1}R_j$ between images $I_i$ and $I_j$. The angle between the two central planes is not determined by the common line. In order to determine it, a third image is added, and the three common line pairs between the three images uniquely determine their relative rotations up to a global reflection. This procedure is known as ``angular reconstitution'' \cite{Goncharov1986,VanHeel1987}. Notice that the handedness of the molecule cannot be determined by single particle cryo-EM, because the original three-dimensional object and its reflection give rise to identical sets of projection images with rotations related by the following conjugation, $\tilde{R}_i = J R_i J^{-1}$, with $J = J^{-1} = \operatorname{diag}(1,1,-1)$. For molecules with non-trivial point group symmetry, e.g., cyclic symmetry, there are multiple common lines between pairs of images, and even self-common lines that enable rotation assignment from fewer images \cite{pragier2019common}.

The identification of common lines is sensitive to noise and angular reconstitution fails at low SNR, such as the SNR of individual particle images. Unlike angular reconstitution that uses only three images and assigns the remaining rotations sequentially, a more noise-robust approach would be to use the entire information between all common lines at once, in an attempt to find a set of rotations for all images simultaneously. There are several such procedures \cite{singer2010detecting,singer2011three,shkolnisky2012viewing,wang2013orientation,bandeira2015non}. Due to space limitations, we only briefly explain the semidefinite programming (SDP) relaxation approach \cite{singer2011three}. Let $(x_{ij},y_{ij})$ be a point on the unit circle indicating the location of the common line between images $I_i$ and $I_j$ in the local coordinate system of image $I_i$. Also, let $c_{ij} = (x_{ij}, y_{ij}, 0)^T$. Then, the common-line property implies that $R_i c_{ij} = R_j c_{ji}$. Such a linear equation can be written for every pair of images, resulting an overdetermined system, because the number of equations is $O(n^2)$, whereas the number of variables associated with the unknown rotations is only $O(n)$. The least squares estimator is the solution to minimization problem
\begin{equation}
\min_{R_1,R_2,\ldots,R_n\in SO(3)} \sum_{i\neq j}\|R_i c_{ij} - R_j c_{ji} \|^2.
\end{equation}
This is a non-convex optimization problem over an exponentially large search space. The SDP relaxation and its rounding procedure are similar in spirit to the Goemans-Williamson SDP approximation algorithm for Max-Cut \cite{goemans1995improved}. Specifically, it consists of optimizing over a set of positive definite matrices with entries related to the rotation ratios $R_i^T R_j$ and satisfying the block diagonal constraints $R_i^T R_i = I$, while relaxing the non-convex rank constraint of this matrix.

Notice that common-line approaches are fully ab-initio in the sense that no initial guess for the 3D structure is ever needed. In fact, common-line approaches estimate rotations rather than a density map. After estimation of rotations the map is determined by classical tomographic reconstruction techniques. Although modern common-line approaches are much more successful than angular reconstitution, estimating rotations directly from experimental images is still challenging due to the high level of noise. Common-line approaches are very effective when applied to 2D class averages \cite{greenberg2017common}.

\subsection{2D classification and averaging}
\label{sec:2D}

\begin{figure}
\includegraphics[width=1.0\textwidth]{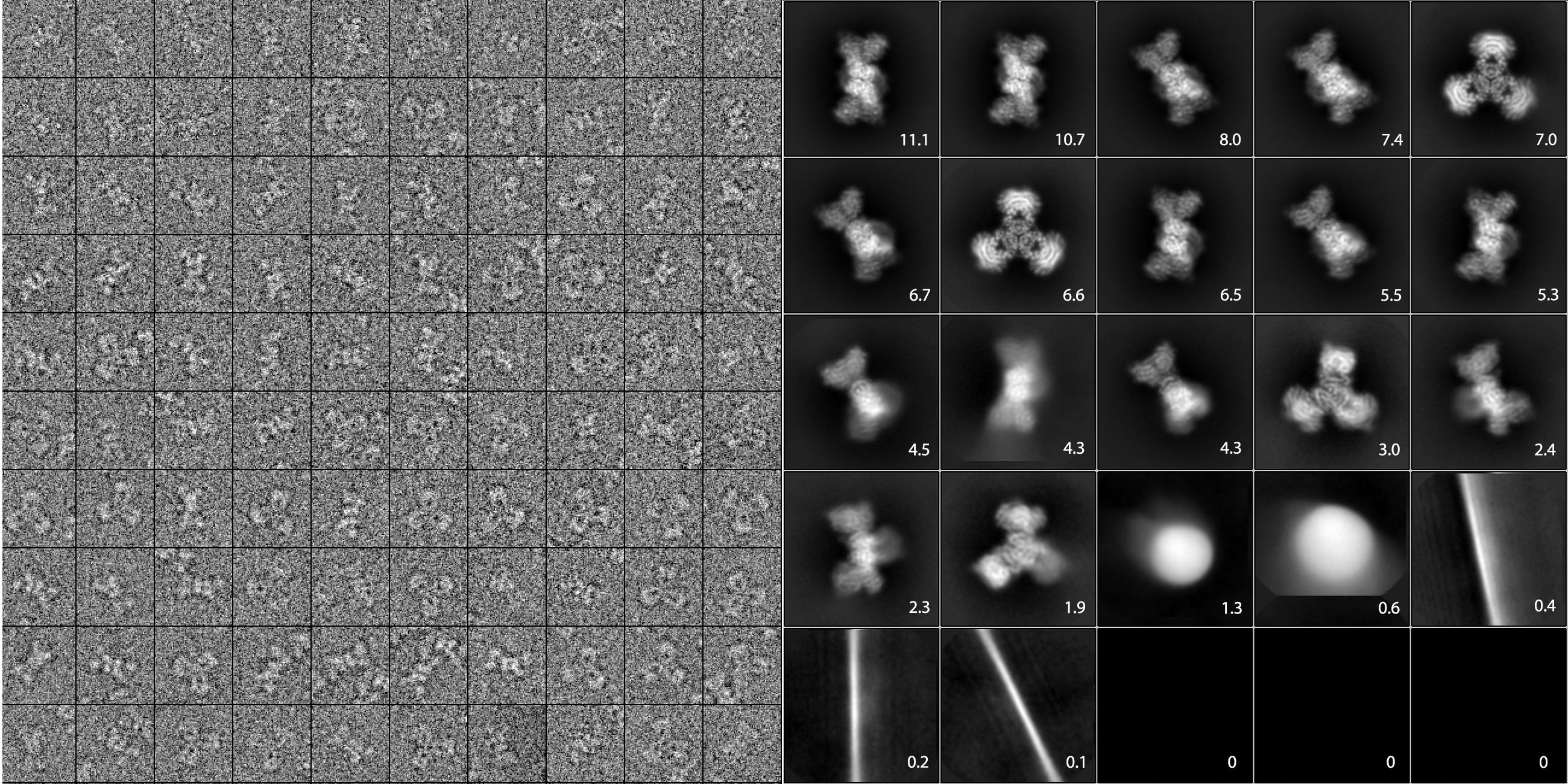}
\caption{Particle images and 2D class averages. {\bf Left}, a gallery of particle images extracted from micrographs. The experimental images are shown cropped to boxes of 256 pixels of $1.06 \rm{\AA}$ per pixel, and contrast has been reversed to make protein ``white''. {\bf Right}, a complete set of 2D class averages computed from 118,000 particles using ML-EM in Relion 2.0. Numbers give the percentage of the total particle set contributing to each average image. Of the eight smallest classes, five come from small numbers of artifact images, and three converged to zero in the ML-EM iterations.}
\label{fig3}
\end{figure}

If images corresponding to similar viewing directions can be identified, they can then be rotationally (and translationally) aligned and averaged to produce ``2D class averages'' that enjoy a higher SNR. The 2D class averages can be used as input to common-line based approaches for rotation assignment, as templates in semi-automatic procedures for particle picking, for symmetry detection, and to provide a quick assessment of the particles and the quality of the collected data. An example of 2D class averages is shown in \textbf{Figure \ref{fig3}}.

There are several computational challenges associated with the 2D classification problem. First, due to the low SNR, it is difficult to detect neighboring images in terms of their viewing directions. It is also not obvious what metric should be used to compare images. Another difficulty is associated with the computational complexity of comparing all pairs of images and finding their optimal in-plane alignment, especially for large datasets consisting of hundreds of thousands of particle images.

PCA of the images offers an efficient way to reduce the dimensionality of the images and is often used in 2D classification procedures \cite{van1981use,van1984multivariate}. Since particle images are just as likely to appear in any in-plane rotation (e.g., by rotating the detector), it makes sense to perform PCA for all images and their uniformly distributed in-plane rotations. The resulting covariance matrix commutes with the group action of in-plane rotation. Therefore, it is block-diagonal in any steerable basis of functions in the form of outer products of radial functions and Fourier angular modes. The resulting procedure, called steerable PCA is therefore more efficiently computed compared to standard PCA \cite{zhao2016fast}. In addition, the block diagonal structure implies a considerable reduction in dimensionality: for images of size $L\times L$, the largest block size is $O(L\times L)$, whereas the original covariance is of size $L^2 \times L^2$. Using results from the spiked covariance model in high dimensional PCA \cite{johnstone}, this implies that the principal components and their eigenvalues are better estimated using steerable PCA, and modern eigenvalue shrinkage procedures can be applied for image denoising and for improved estimation of the covariance matrix of the 2D images \cite{bhamre2016denoising}.

The steerable PCA framework also paves the way to a natural rotational invariant representation of the images using the bispectrum \cite{zhao2014rotationally}. Images can therefore be compared using their rotational invariant representation, saving the cost associated with rotational alignment. In addition, efficient algorithms for approximate nearest neighbors search can be applied for initial classification of the images. The classification can be further improved by applying vector diffusion maps \cite{singer2012vector,singer2011viewing}, a non-linear dimensionality reduction method that generalizes Laplacian eigenmaps \cite{belkin2002laplacian} and diffusion maps \cite{lafon} by also exploiting the optimal in-plane transformation between neighboring images.

A popular way for 2D classification and averaging is using a maximum likelihood framework similar to that of 3D reconstruction and classification \cite{scheres2005maximum}. Specifically, each experimental image is assumed to be formed from one of $K$ fixed images followed by in-plane rotation, translation, corruption by CTF and noise. The in-plane rotations, translations, and class assignments are treated as latent variables, and expectation-maximization is used to find the maximum likelihood estimator of the $K$ class averages. Note that since viewing directions are arbitrarily distributed over the sphere, the statistical model with a finite $K$ is inaccurate. A large value of $K$ leads however to high computational complexity and to low-quality results because few images are assigned to each class. ML-EM also suffers from the ``rich getting richer'' phenomenon: most experimental images would correlate well with, and thus be assigned to the class averages
that enjoy higher SNR. As a result, ML-EM tends to output only a few, low-resolution classes. This phenomenon was recognized in \cite{sorzano2010clustering} and is also present for ML-EM based 3D
classification and refinement.

All formulations for 2D classification completely ignore the image formation model, namely that images are tomographic projection images of a fixed 3D density. Instead, 2D classification is regarded as an unsupervised learning problem for which a variety of clustering approaches can be applied. It is therefore worth keeping in mind that for some challenging datasets 2D classification may fail to produce high-quality class averages, yet 3D reconstruction may still prevail, as the latter makes use of the image formation model.

\subsection{Method of moments}

The method of moments is a classical way to estimate model parameters from data moments going back to the 19th century. In fact it predates maximum likelihood estimation that was introduced in the 20th century. The method of moments for SPR was first proposed by Zvi Kam \cite{kam1980}, again preceding the maximum likelihood approaches in the field \cite{sigworth1998maximum}. The method of moments for SPR is based on relating low-order statistics of the 2D projection images with the unknown 3D density map without any attempt to estimate the image orientations in the process.

The first moment is simply the sample average of the projection images. Due to in-plane rotation invariance, the level sets of the averaged image are concentric circles and the image only depends on the radial distance. The second moment is closely related to the covariance matrix of the 2D images; the two only differ by a term that depends on the first moment. The covariance matrix is of size $L^2 \times L^2$, but its block-diagonal structure (see Section \ref{sec:2D}) implies that it has only $O(L^3)$ non-zero entries. In principle that might be sufficient to uniquely determine an $L\times L \times L$ 3D density map. However, in the case of uniform distribution of viewing directions, Kam invoked the Fourier projection slice theorem to show that the information provided by the covariance matrix is equivalent to the autocorrelation over the rotation group SO(3) of the Fourier transformed density map with itself. For that reason, Kam's theory is often referred to as Kam's autocorrelation analysis. The 3D density map cannot be uniquely determined from the autocorrelation function. Instead, when expanding the map in a basis of spherical harmonics and radial functions, the expansion is determined up to a series of missing orthogonal matrices. The missing orthogonal matrices can be solved from the third order moment \cite{bandeira2017estimation}.

It is worth noting that Kam's theory has so far received little attention in the cryo-EM community. It is an idea that was clearly ahead of its time: there was simply not enough data to accurately estimate second and third order statistics from the small datasets that were available at the time (e.g. typically just dozens of particles). Moreover, accurate estimation of such statistics requires modern techniques from high dimensional statistical analysis that have only been introduced in the 21st century. Estimation is also challenging due to the varying CTF between micrographs and non-perfect centering of the images. Most importantly, Kam's method requires a uniform distribution of particle orientations in the sample, an assumption that usually does not hold in practice, and severely limits its application. In an effort to make Kam's theory a viable approach for 3D ab-initio modeling, it has been recently extended and generalized in several ways to meet the challenges of non-perfect centering, CTF, high dimensional statistical estimation, and non-uniform distribution of viewing directions \cite{Bhamre2015,bhamre2016denoising,levin20183d,sharon2019method}.

The main advantage of the method of moments over maximum likelihood and Bayesian inference approaches is that it requires just one pass over the data for the computation of the moment statistics. In fact, this computation can be done in a streaming, on-the-fly fashion and can be naturally parallelized. The recovery of the 3D density from the moment statistics is a challenging optimization problem that requires solving a large system of polynomial equations. As the computational complexity and conditioning scale badly with $L$, the method of moments is envisioned to shine for low-resolution ab-initio modeling, but is not expected to be the method of choice for high-resolution reconstruction. Also from the statistical estimation viewpoint, the method of moments is not an efficient estimator and therefore cannot achieve the same resolution as maximum likelihood. Another interesting future application of the method of moments is 3D reconstruction without particle picking \cite{bendory2018toward}, which could be useful for reconstructing particles that are too small to be reliably detected.

\section{HETEROGENEITY ANALYSIS}

Unlike X-ray crystallography and NMR that measure ensembles of particles,
single particle cryo-EM produces images of individual particles. Cryo-EM therefore has the potential to analyze compositionally and conformationally heterogeneous mixtures and, consequently, can be used to determine the structures of complexes in different functional states.

The heterogeneity problem of mapping the structural variability of macromolecules is widely recognized nowadays as the main computational challenge in cryo-EM \cite{nogales2016development,glaeser2016good,subramaniam2016cryoem,sorzano2017challenges}. Much progress has been made in distinguishing between a small number of distinct conformations, but the reconstruction of highly similar states and ``3D movies'' of the continuum of states remain an elusive unfulfilled promise.

Indeed, current software packages offer tools for 3D classification (or 3D sorting) and reconstruction of a small number of distinct conformations that are typically based on the maximum likelihood and Bayesian inference frameworks used for 3D reconstruction \cite{scheres2007disentangling,scheres-relion,scheres2016processing,brubaker,ludtke2016single,de2013xmipp,lyumkis2013likelihood,grant2018cistem}. Specifically, the parameter of interest in this case are $K$ different structures $\phi_1,\ldots,\phi_K$ (with $K > 1$), and for each image the latent variables now also include the class assignment. These methods for 3D classification are quite powerful when $K$ is small. However, the quality of their output deteriorates significantly as $K$ increases, because for large $K$  each class is assigned a small number of images and class assignment therefore becomes more challenging. Moreover, the computational complexity of 3D classification increases linearly with $K$. A common practice is to run 3D classification with a small $K$, discard bad classes and their corresponding images, and perform hierarchical classification on the good classes (typically using masking). There is a great deal of manual tweaking that puts in question the reliability and reproducibility of the analysis. There is a wide consensus that more computational tools are needed for analyzing highly mobile biomolecules with many, in particular, continuous spectrum of conformational changes, and method development for analyzing continuous variability is a very active field of research \cite{sorzano2019survey}.

One useful approach for characterizing the structural variability is by estimating the variance-covariance matrix of the 3D structures in the sample. The basic theory of \cite{liu1995estimation-a,liu1995estimation-b} has been considerably developed over the years \cite{penczek2002variance,penczek2006estimation,zhang2008heterogeneity,penczek-pca,frank70s_10k,zheng2012three,wang2013dynamics,xu2018allosteric,tagare,klaholz2015structure,gene,isbi15,anden2018structural}.
The covariance matrix of the population of 3D structures can be estimated using the Fourier slice theorem for covariance. Specifically, for every pair of non-collinear frequencies, together with the origin they uniquely define a central slice; therefore, the covariance can be computed for any pair of frequencies given a full coverage of viewing directions. In contrast to classical 3D reconstruction which is well-posed even with partial coverage of viewing directions (as long as each frequency is covered by a slice), covariance estimation requires full coverage which is a far more stringent requirement. Moreover, the covariance matrix is of size $L^3 \times L^3$, and storing $O(L^6)$ entries increases quickly with $L$. The computational complexity and the number of images required for accurate estimation also grow quickly as a function of $L$. As a result, the covariance matrix is usually estimated only at low-resolution.  Still, the variance map indicates flexible regions of the molecule, and the leading principal components of the covariance matrix offer considerable dimension reduction.

Methods based on normal mode analysis of the 3D reconstruction \cite{jin2014iterative,sorzano2014hybrid,jonic2017computational} or of the multiple 3D classes \cite{schilbach2017structures} have also been applied. However, extrapolating the harmonic oscillator approximation underlying normal mode analysis to complex elastic deformations is questionable.

Methods based on manifold learning, specifically, diffusion maps, have also been proposed to analyze continuous variability. Similar to the covariance matrix approach, also here the rotations are assumed to be known. One approach performs manifold learning on disjoint subsets of 2D images corresponding to similar viewing directions \cite{dashti,frank2016manifold,dashti2018functional}. In this way, the heterogeneity at the level of 2D images can be characterized, and insight into energy landscapes can be obtained. However, the nature of 3D variability is not fully resolved because stitching the manifolds resulting from different viewing directions does not have an adequate solution at present. An alternative diffusion map approach that performs directly at the 3D level has been recently developed, but its evaluation on experimental datasets in still ongoing \cite{moscovich2019cryo}.

Methods for continuous heterogeneity analysis that do not require known rotations have also been recently introduced. Multi-body refinement \cite{shan2016local,nakane2018characterisation} takes a segmentation of a 3D molecular reconstruction and attempts to refine each part separately from a static base model, with independent viewing directions and shift parameters for each part. Multi-body refinement, however, may fail to accurately reconstruct the interface between moving parts and it cannot capture non-rigid movements typical of complex macromolecules. Other recent approaches that are still under development and aim to recover continuous heterogeneity jointly with rotations are based on representing the 3D movie of molecules as a higher dimensional function \cite{lederman2019hyper} and using deep neural nets \cite{zhong2019reconstructing}. The latter builds on a spatial variational autoencoder introduced in \cite{bepler2019explicitly}.

\section{EVALUATING THE RESULTS}
\subsection{Resolution measures}

In X-ray crystallography the quality of the raw data is apparent from the pattern of diffraction spots--if they extend to high $k$, then the eventual 3D density map is likely to have high resolution. In the case of single-particle reconstruction the quality of data is not so apparent. Even the best single particle images have an SNR well below unity for spatial frequencies higher than 0.05 or 0.1 $\rm{\AA}^{-1}$, so how can the quality of the reconstruction be estimated? The current methods are based on an evaluation of the final 3D map. One approach is to estimate a correlation coefficient between the high-frequency components of two maps constructed from the same dataset. A second approach is to ask, is the signal in the map significantly different from noise.

\begin{figure}
\includegraphics[width=1.0\textwidth]{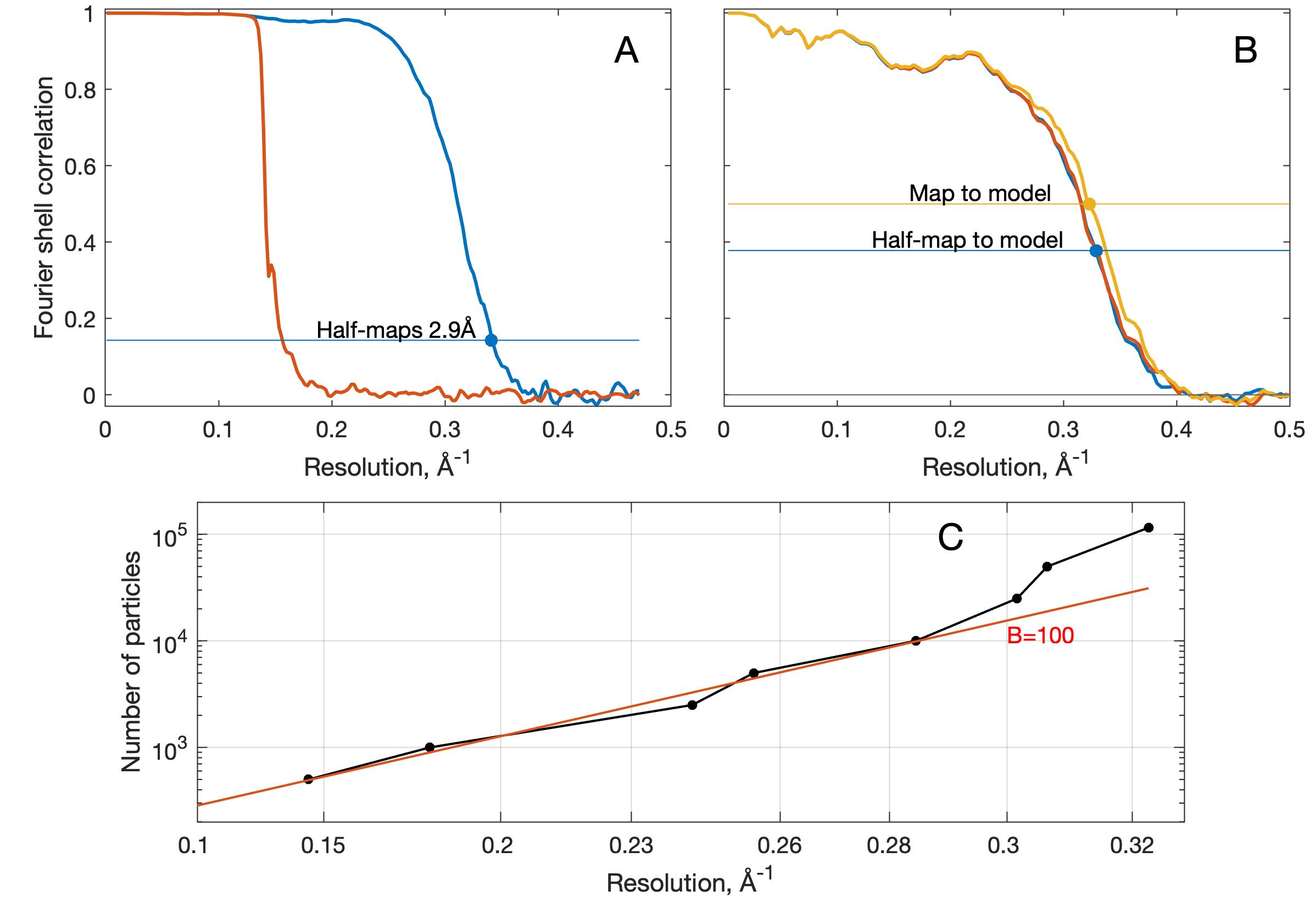}
\caption{Resolution estimation for the PaaZ reconstruction. {\bf A}, the blue curve shows the Fourier shell correlation between masked half-maps (3D reconstructions from disjoint halves of the dataset). Using the criterion of $\text{FSC}=0.143$, the resolution of a reconstruction from the full dataset is $2.9 \rm{\AA}$.  The red curve is the FSC where in each of the half-maps the phases are randomized beyond $7 \rm{\AA}$ resolution. The lack of correlations above that spatial frequency shows that masking is not introducing artifacts in the FSC. {\bf B}, Fourier shell correlations between each of the half-maps and a density calculated from the atomic model
(red and blue curves, nearly indistinguishable).
As the model was built using only one half-map, the close agreement is consistent with the absence of overfitting. The FSC between the full map and the model is also shown as the yellow curve. Using the criteria of $\text{FSC}=\sqrt{0.143}$ and $\rm{FSC}=0.5$, respectively, a resolution estimate of $3.0 \rm{\AA}$ is obtained in each case. {\bf C}, a Wilson plot of resolutions obtained from varous-sized subsets of particle images in a PaaZ dataset. The plot of $\ln N$ vs. $k_0^2$ yields a straight line corresponding to
$B=100 \rm{\AA}^{-2}$ for all but the largest datasets.}
\label{fig4}
\end{figure}

The first of these methods employs the Fourier Shell Correlation (FSC), which computes a correlation coefficient over a spherical shell between two 3D Fourier-space maps \cite{harauz1986exact} (see \textbf{Figure 4A}). It is defined as
\begin{equation}
\text{FSC}(k)=\frac{\sum_{s \in S_k}{U_s V^*_s}} {\sqrt{(\sum_{s \in S_k}{|U_s|^2})(\sum_{s \in S_k}{|U_s|^2})}}
\end{equation}
where $S_k$ is the set of Fourier voxels in a spherical shell at distance $k$ from the origin, and $U$ and $V$ are the Fourier transforms of the maps $u$ and $v$ to be compared. To estimate the resolution of a reconstruction made from a given dataset, that dataset is divided randomly into two halves, and two separate 3D reconstructions are made. The FSC is then computed between these two ``half maps''. Noise will show no correlation between the two half maps as they are made from disjoint sets of data, while any true signal will yield a correlation. The FSC is typically close to 1 at small $k$, as low-frequency signal is strong, but decays to zero at high $k$ as the signal is lost in noise. By convention the resolution of the full map is taken to be the spatial frequency where the half-map FSC falls to the value 0.143, a criterion that is chosen to match resolution criteria used in X-ray crystallography \cite{rosenthal2003optimal}. Let the Fourier-transformed maps be written $U=X+N_U$ and $V=X+N_V$ where $X$ is the true signal, assumed to be identical in the two half-maps, while $N_U$ and $N_V$ are independent noise terms for which the expectation $P^2$ of the power spectrum is the same. Then the expectation of the FSC is

\begin{equation}
\langle \text{FSC}(k) \rangle =\frac{\sum_{s \in S_k}{X_sX_s^*}} {\sum_{s \in S_k}{X_sX_s^*}+\sum_{s \in S_k}P^2_s},
\end{equation}
which is seen to be closely related to the spectral signal-to-noise ratio for each of the half-maps,
\begin{equation}
\langle \text{SSNR}(k) \rangle =\frac{\sum_{s \in S_k}{X_sX_s^*}} {\sum_{s \in S_k}P^2_s}.
\end{equation}
In the derivation of the 0.143 criterion, this is assumed to be half of the SSNR for a map created from the entire dataset.

Another way the FSC is used is to compare a cryo-EM map with an atomic model; this is called the map-to-model FSC (\textbf{Figure 4B}) where the model is illustrated in \textbf{Figure 5}. An artificial density map is created from atomic coordinates, for example by placing point-densities at the atom positions, and the FSC is computed. The resolution of the cryo-EM map is then taken to be the value of $k$ where the FSC falls to 0.5. This criterion matches the 0.143 criterion for the comparison of half-maps \cite{rosenthal2003optimal}, and corresponds to the point at which the SSNR of the full map falls to 1/3.

When one obtains a map from single-particle reconstruction, the size of the reconstruction box can be chosen arbitrarily, and in many cases the support of the particle is a small fraction of the volume of the box. A problem is that FSC values are lower when the volume of the region outside the particle, containing only noise, is increased \cite{sindelar2012optimal}. Users therefore apply a mask to the reconstruction, where voxels outside the particle support are set to zero, and one is tempted to use very tight masks to maximize the FSC values. There is however a danger of inflated FSC values with sharp masks, as will be discussed in Section 7.3 below.

Due to flexibility or heterogeneity in the particles of the dataset, it is usually the case that some parts of the 3D map have lower resolution than other parts, as fine details are blurred. An approach to determine the local resolution is to compute the FSC in small volumes, using appropriate window functions and taking into account the effect of small windows on the details of the FSC \cite{cardone2013one}.

Perhaps an extreme case of windowed Fourier transforms is the use of small, steerable filter kernels \cite{kucukelbir2014quantifying} to obtain an isotropic resolution measure for each voxel. In this case the criterion for establishing the resolution is not the correlation between two maps, as in FSC, but a statistical test for the presence of signal over the noise level.

In conventional microscopy resolution is defined, for example in the Rayleigh criterion, as the minimum spacing between a pair of point-scattering objects where they become distinguishable. None of the approaches discussed above evaluate the resolution of the imaging system in this sense \cite{penczek2010resolution}.  Instead as discussed above they are based on measures of spectral signal-to-noise ratio. However, once a researcher determines a resolution value for a map in one of these ways, the next step is to apply a lowpass filter at that spatial frequency with the goal of suppressing high-frequency details that arise from noise. The final result is that the map contains Fourier components only up to the resolution limit, so that the Rayleigh criterion is, in the end, approximately satisfied due to the lowpass filtering.

When there is particle flexibility or when there is a bias in viewing directions, the resolution of a map is anisotropic. Measures of resolution anisotropy have recently been proposed \cite{tan2017addressing}.

\subsection{B-factor and particle number}

The power spectrum of the signal in a map is remarkably well described by Wilson statistics \cite{wilson1942determination} where the macromolecule is modeled as a bag of randomly-positioned atoms. Modeling atoms as point densities, the power spectrum of a map is expected to be constant at high resolutions, in practice above about $0.1 \rm{\AA}^{-1}$. There is a low-resolution excess in the spectral density that represents the shape of the bag, because the average electron-scattering density in the macromolecule differs from the average density of the surrounding ice.

Because of the rolloff of high frequency information in experimental images, modeled as a  envelope function, then in the case of uniformly distributed projection directions the reconstructed volume will show a Gaussian rolloff with the same factor $B$. The resolution of a reconstruction, evaluated as the frequency $k_0$ at which the ${\rm SSNR}(k_0)= 1/3$, nevertheless can increase without limit, albeit slowly, as the number of particles increases. This is because the coherently-adding signal, even at frequencies where the Gaussian rolloff is severe, increases to become comparable to the noise spectral density. At high frequencies $ k>~ 0.1 \rm{\AA}^{-1} $ we can make the Wilson assumption of constant signal power in an ideal reconstruction, and also assume that the noise spectral density is constant, so that the spectral signal to noise ratio will be
\begin{equation}
	{\rm SSNR}(k) = Nc e^{-B{k}^2/2}
\end{equation}
where $c$ is a constant. If we set the resolution $k_0$ to satisfy ${\rm SSNR}(k_0)=1/3$ then $k_0$ depends on $B$ and $N$ as
\begin{equation}
	k_0^2= 2 \ln (3Nc)/B
\end{equation}

The overall quality of the dataset can therefore be evaluated by observing the dependence of $k_0$ on $N$ as subsets of $N$ particle images are used in separate reconstructions. \textbf{Figure 4C} shows a plot of $k_0^2$ as a function of $\operatorname{ln} N$, a so-called Wilson plot. In this example, subsets of $10^3$ to $10^4$ particles yielded resolutions well described by this relation with $B=100 \rm{\AA}^2$. Some unknown effect seemed to reduce the resolution from larger $N$.

\begin{figure}
\includegraphics[width=1.0\textwidth]{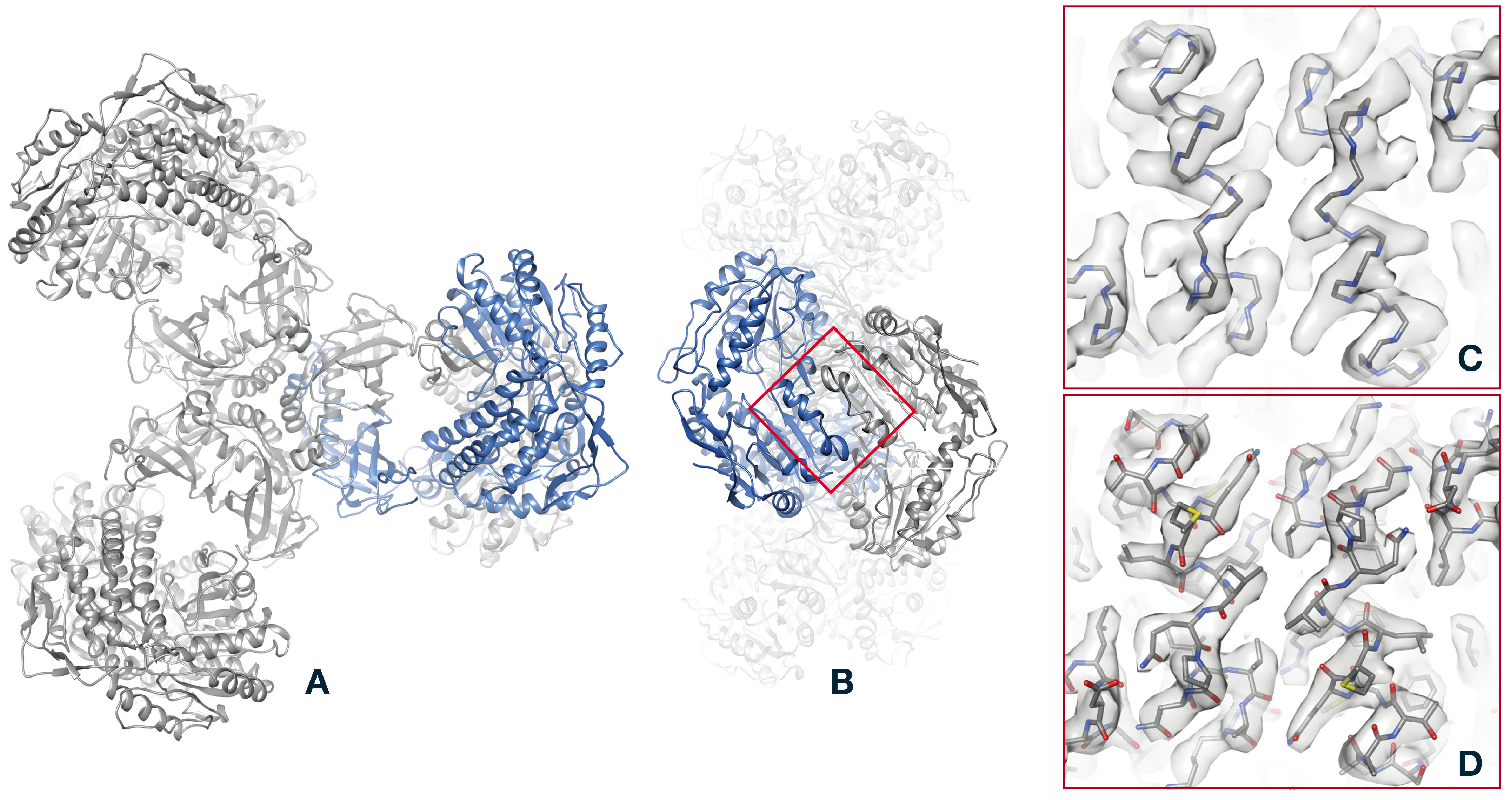}
\caption{The last step in structure determination: fit of an atomic model to the PaaZ reconstruction. {\bf A}, overview of the PaaZ model, with ribbons denoting the folding of the polypeptide chain. To give an idea of the physical scale of the model, the helical structures (alpha helices) have a pitch of $5.4 \rm{\AA}$. One of the six identical subunits in the PaaZ macromolecule is colored blue. {\bf B}, an ``end'' view of the model, obtained by a $90^\circ$ rotation. {\bf C}, superposition of the polypeptide backbone on the density map (gray surface rendering), shown for the region of two alpha helices marked by the red square in B. {\bf D}, comparison of the map and model, this time with all non-hydrogen atoms shown. Although individual atoms are not resolved in the $2.9 \rm{\AA}$ map, bulges in the helical features of the map constrain the angles of flexible bonds in the amino-acid side chains, so that atom positions are well determined.}
\label{fig5}
\end{figure}

\subsection{Validation of results}

Because the SNR of particle images is so low, it is not possible to evaluate the quality of a dataset from the raw images. Further, there is considerable danger of obtaining false results from poor data, for example from the  ``Einstein from noise'' effect, see Section \ref{sec:picking}. In view of problems like this, several approaches have been suggested to validate results.

Artifactually high half-map FSC values result from any commonality between the two half maps. It is now standard practice to use strongly lowpass-filtered references in reference-based particle picking as well as a strongly filtered initial map for 3D refinement. Both of these measures are meant to remove any common inputs to the refinement process except at very low spatial frequencies, for example $k< .025 \rm{\AA}^{-1}$. Starting with the initial map, the refinements of the half-maps are carried out completely independently, the only further interaction being a translational and rotational alignment of the half-maps for the computation of the FSC. To minimize artifacts from that step, the alignment is also carried out between strongly lowpass-filtered volumes. Finally, artifactually large FSC values can result when a common real-space mask with sharp edges is used to multiply each half-map, for example to exclude noise from outside the molecular envelope. A test \cite{chen2013high} for an artifact of this sort makes use of modifying the particle images by randomizing the phases of Fourier components at high frequencies (see \textbf{Figure 4A}). The expectation is then that the resulting half-volumes should have no correlation at these frequencies.

Especially before the widespread use of maximum-likelihood refinement, another pitfall was the assignment of incorrect particle orientations. To test for this, pairs of micrographs are acquired, one as usual with no tilt of the specimen, and a second with the specimen tilted at a small angle of $10^\circ$ or $20^\circ$. For each particle common to the two micrographs, the orientations should differ by the tilt angle, and deviations from this relationship imply errors in angle assignment \cite{henderson2011tilt}.

Finally there is a possibility of errors in building the atomic model, if noise in the map results in the mis-assignment of atom positions. In X-ray crystallography it was realized that a low rms deviation between map and model might simply reflect over-fitting of a model. To avoid this problem, it is now common practice to withhold a fraction, for example 10\% of the X-ray data from the refinement process, and then compute the rms errors between those data and the final model. The result is called the $R_{\rm{free}}$ value \cite{brunger1992free}. In cryo-EM there is not a simple way to perform this sort of test; instead a rigorous approach is to build the atomic model using only one half map, and then compute the map-to-model FSC using the other half map, as was done with the PaaZ model (\textbf{Figure \ref{fig4}B}).

\subsection{Asymptotic behavior}

Deviations between half maps can come from two sources. First, in the Wilson plot and other interpretations of the FSC, deviations are assumed to come from noise in the individual particle images, while the assumption is made that orientation assignments are as precise in the reconstruction of a half map as in a full map. This seems to be a vaild assumption for most high-resolution work \cite{scheres2012prevention}, and overall SNR is proportional to the particle number $N$ used in deriving a map.  However, this will not be the case if the particle image SNR is particulaly low. In this case a reduced $N$ yields a noisier 3D map which in turn produces larger errors in angle assignments.

In \cite{sigworth1998maximum} the ML-EM algorithm was applied to a toy problem in 2D alignment of images. In datasets of 200,000 simulated particle images the noise variance was varied to produce a variety of image SNR values. When the image SNR was above a certain threshold, where translation and rotation assignments were reliable, the SNR of the reconstructed average image increased proportionately. However with the image SNR below the threshold the probability of correct alignment dropped off rapidly, and the output SNR was seen to be more sensitive to the input SNR, with an apparent third power relationship. This may be the case in studies of particularly small particles, such as myoglobin. In a study of hemoglobin (molecular weight 64 kDa) using a microscope with a Volta phase plate, a dataset of 172,000 particles yielded a resolution of $3.2  \rm{\AA}$ resolution, while in the case of myoglobin (17 kDa) and presumably a similar-sized dataset the reconstruction remained at a much lower resolution \cite{khoshouei2017revisiting}.

The mathematical model of multi-reference alignment provides a rigorous theoretical explanation for this phenomenon \cite{abbe2018multireference,abbe2018estimation,perry2019sample,ma2018heterogeneous,bandeira2017estimation}. Specifically, the theory suggests that there are two different regimes for the SNR: At high SNR, the estimation error (MSE) is inversely proportional to the SNR (as if the latent variables are estimated accurately), while at low SNR the MSE is inversely proportional to $\text{SNR}^d$, where $d$ is the ``cut-off'' moment, that is, the lowest order moment (in the method of moments) that distinguishes between any two generic signals. For example, $d=3$ for multi-reference alignment of images with uniform distribution of in-plane rotations, but $d=2$ if the distribution in non-uniform (or, more precisely, aperiodic).

\section{FUTURE CHALLENGES}

\subsection{Specimen preparation and protein denaturation}

A protein macromolecule is a linear polymer of amino acids that folds into a particular shape due to various non-covalent interactions. The surface of a protein has many polar and charged groups that interact strongly with the surrounding water molecules, while the interior is hydrophobic, with essentially all polar groups stabilized by fixed hydrogen bonds. When exposed to a hydrophobic environment or an air-water interface most proteins unfold--that is, denature--bringing the hydrophobic interior residues to the surface. Exposed hydrophobic regions in neighboring proteins coalesce, yielding amorphous aggregates of misfolded proteins.

Recall that a cryo-EM specimen consists of a film of protein solution only a few hundred angstroms thick which is quickly frozen to produce an amorphous ice layer. During the time before freezing, typically several seconds, a protein molecule diffuses rapidly (diffusion constant $ \sim 10^{13} \rm{\AA ^2 /s} $) and has many opportunities to diffuse up to the edge of the film. Some proteins appear to remain as intact single particles in cryo-EM specimens, while others form large aggregates that are useless for structure determination. The best remedy appears to be the provision of solid substrate to anchor proteins within the aqueous layer. Such an aggregation problem was encountered in the PaaZ study \cite{sathyanarayanan2019molecular} and was solved by the provision of a layer of graphene oxide bounding the aqueous film. The protein adsorbs to the graphene oxide and is thereby protected from the air-water interface. Graphene and its derivatives form an ideal support film because graphene is a two-dimensional crystal only one carbon-atom thick, and so introduces very little electron scattering to the specimen. The fabrication of EM grids with graphene derivatives of good surface properties is unfortunately still an art form, such that a general substrate for single particle work remains to be developed.

Another problem that arises in specimen preparation is preferred orientation of particles. Many proteins orient with one surface preferentially interacting with the air-water interface, or preferentially interacting with a supporting film. The result is an uneven distribution of projection angles, resulting in poor 3D reconstructions. Sometimes the introduction of surfactants, changes in pH or other measures can mitigate the preferred orientations. If necessary, data are collected from tilted specimens \cite{tan2017addressing} because this provides a larger range of orientations. However, tilting the sample is disadvantageous for high-resolution work because the effects of beam-induced specimen motion are increased. This is because the largest motions are normal to the specimen plane. New computational approaches for datasets with preferred orientations would be very welcome in the field.

One solution to both of these issues in cryo-EM sample preparation would be to image the macromolecules without removing them from their native cellular environment. This is commonly done (but with low-resolution results) in cryo-EM tomography. Might this also be possible for SPR, given better computational tools? The very crowded environment inside cells results in greatly increased background noise, but promising results in the detection and orientation of large macromolecules (ribosomes) have been reported from cellular imaging \cite{rickgauer2017single}.

\subsection{The $B$ factor and resolution limitation}

The envelope function of the CTF, summarized by the $B$ factor, is the ultimate limitation in high-resolution cryo-EM projects. At present the $B$-factor from a high-quality dataset is roughly 100, which as we saw in Section \ref{sec:phase} represents an attenuation of signal to $1/e$ already at $5 \rm{\AA}$ resolution, or $1/e^2$ at $2.5 \rm{\AA}$. Physical contributions to this limitation in resolution have been described in a recent review \cite{glaeser2019good} and are summarized here.

Current high-end microscopes have highly coherent electron beams and extremely stable stages, so they make little contribution to the blurring represented by the $B$-factor. Charging of the specimen also makes little contribution if proper imaging conditions are used \cite{russo2018microscopic}. There is however unavoidable particle movement due to beam-induced rearrangement of water molecules in the vitreous ice \cite{mcmullan2015thon}. Water molecules move with a diffusion constant of about $1 \rm{\AA}^2$ per electron dose of $1 e/\rm{\AA}^2$ and so produce Brownian motion of particles. This motion may become substantial for smaller particles, and per-particle motion correction has been implemented \cite{zivanov2019bayesian}.

The largest contribution to the $B$-factor seems to be uncorrected beam-induced specimen movement. Motion correction during exposure has greatly improved cryo-EM imaging, but it is during the initial dose of $1-2 e/ \rm{\AA}^2$, typically captured in the first frame of an acquired movie, that the largest specimen movements occur. Even if a faster movie frame rate is employed it is not clear that there is sufficient signal to allow these movements to be adequately tracked and compensated, but improvements from novel computational approaches would have a large effect. Radiation damage is typically characterized as a loss of high-resolution signal, and after $2 e/\rm{\AA}$ of exposure the radiation damage corresponds to more than half of the $B$-factor value that is observed.

\subsection{Autonomous data processing}

Traditionally there is user intervention in two parts of the data processing pipeline. At the stage of particle picking, the user checks the particle selection and tunes parameters, such as the selection threshold, to ensure a low number of false-positives. Farther along the processing pipeline, users perform 2D classification or reconstruct multiple 3D volumes (3D classification) to further select a subset of the particle images that produce the "best" reconstruction. It often occurs that a majority of particle images are rejected at this stage.

Researchers are often happy to reject large portions of their datasets in order to achieve higher-resolution reconstructions.  The rejected particles are sometimes contaminating macromolecules, or complexes missing one or more subunits. Rejected particles may be partially denatured or are aggregated. Improvements in biochemical purity and in specimen preparation may reduce these populations in the future, but meanwhile improved outlier-rejection algorithms would help greatly. However there is also the possibility that rejected particles represent alternative conformations of the complex, or states in which there is more flexibility; in these cases a more objective process of evaluating particle images through an entirely automated pipeline would be very desirable.

Another reason for automation of the processing pipeline has to do with quality control of data acquisition. In a cryo-EM structure project traditionally one spends some hours or days collecting micrographs, and then carries out the image processing offline over some days or weeks afterward. Unfortunately only after creation of the final 3D map does one know for sure that a dataset yields a high-resolution structure. Some hints of data quality can be obtained along the way: correlations to high resolution in power spectrum fits (\textbf{Figure \ref{fig1}D}) and the visibility of fine structure in 2D classes provide some confidence that a high-resolution structure will result. Automatic data processing, concurrent with acquisition, is currently being implemented at cryo-EM centers. This is beginning to allow experimenters to evaluate quickly the data quality at the 2D classification stage and even with 3D maps. Such real-time evaluation can greatly increase the throughput of cryo-EM structure determination. Most importantly it accelerates the tedious process of specimen optimization. Further, it allows for much better use of time on expensive microscopes for high-resolution data collection.

\section*{DISCLOSURE STATEMENT}

The authors are not aware of any affiliations, memberships, funding, or financial holdings that might be perceived as affecting the objectivity of this review.

\section*{ACKNOWLEDGMENTS}

A.S. was partially supported by NSF BIGDATA Award IIS-1837992, award FA9550-17-1-0291 from AFOSR, the Simons Foundation Math+X Investigator Award, and the Moore Foundation Data-Driven Discovery Investigator Award. F.J.S. was supported by NIH grants R25 EY029125 and R01NS021501. The authors are grateful to G. Cannone (MRC Laboratory of Molecular Biology, Cambridge, U.K.) and K.R. Vinothkumar (National Center for Biological Sciences, Bangalore, India) for sharing the raw images and intermediate data underlying their published study \cite{sathyanarayanan2019molecular}. The authors thank Philip Baldwin, Alberto Bartesaghi, Tamir Bendory, Nicolas Boumal, Ayelet Heimowitz, Ti-Yen Lan, and Amit Moscovich for commenting on an earlier version of this review.

%
%
%
%
%

\bibliographystyle{ar-style3}
\bibliography{refs2019}

\end{document}